\documentclass[a4paper,structabstract]{aa}
\usepackage{amsmath}
\usepackage{amssymb}
\usepackage{graphicx}
\usepackage{subfigure}
\usepackage[innercaption]{sidecap}
\usepackage{epsfig}
\usepackage{color}
\usepackage{txfonts}
\usepackage{natbib}
\usepackage[bookmarks=false,colorlinks=true, citecolor=blue, breaklinks=true]{hyperref} 
\bibpunct{(}{)}{;}{a}{}{,} 

\newcommand{\ds}{\displaystyle}
\newcommand{\ftns}{\footnotesize}
\newcommand{\Msun}{M_{\odot}}
\newcommand{\Ms}{M_{\star}}
\newcommand{\GAS}{\texttt{G.A.S.}}
\newcommand{\kms}{km s$^{-1}$}

\begin{document}

\title{\GAS\ I: A prescription for turbulence-regulated star formation and its impact on galaxy properties} 
\authorrunning{Cousin, Guillard, \& Lehnert}

\titlerunning{Turbulence-regulated star formation in the G.A.S semi-analytical model for galaxy formation}

\author{M. Cousin\inst{1}, P. Guillard\inst{2,3}, \and M.~D. Lehnert\inst{2}}

\institute{Aix Marseille Univ, CNRS, CNES, LAM, Marseille, France.\\ \email{morgane.cousin86@gmail.com}, website: {\tt morganecousin.wordpress.com}
\and Sorbonne Universit\'e, CNRS UMR 7095, Institut d'Astrophysique de Paris, 98 bis bd Arago, 75014 Paris, France
\and
Institut Universitaire de France, Ministère de l'Education Nationale, de l'Enseignement Supérieur et de la Recherche, 1 rue Descartes, 75231 Paris Cedex 05, France
}

\date{Received November 19 2018 / Accepted January 29 2019}

\abstract{Star formation in galaxies is inefficient, and understanding how star formation is regulated in galaxies is one of the most fundamental challenges of contemporary astrophysics. Radiative cooling, feedback from supernovae and active galactic nuclei, large-scale dynamics and dissipation of turbulent energy act over various time and spatial scales, and all regulate star formation in a complex gas cycle.}{This paper presents the physics implemented in a new semi-analytical model of galaxy formation and evolution: \GAS\,.}{The fundamental underpinning of our new model is the development of a multi-phase interstellar medium (ISM) in which energy produced by supernovae and active galaxy nuclei maintains an equilibrium between the diffuse, hot, stable gas and a cooler, clumpy, low-volume filling factor gas. The hot gas is susceptible to thermal and dynamical instabilities. We include a description of how turbulence leads to the formation of giant molecular clouds through an inertial turbulent energy cascade, assuming a constant kinetic energy transfer per unit volume. We explicitly model the evolution of the velocity dispersion at different scales of the cascade and account for thermal instabilities in the hot halo gas. Thermal instabilities effectively reduces the impact of radiative cooling and moderates accretion rates onto galaxies, and in particular, for those residing in massive halos.}{We show that rapid and multiple exchanges between diffuse and unstable gas phases strongly regulates star-formation rates in galaxies because only a small fraction of the unstable gas is forming stars. We checked that the characteristic timescales describing the gas cycle, the gas depletion timescale and the star-forming laws at different scales are in good agreement with observations. For high mass halos and galaxies, cooling is naturally regulated by the growth of thermal instabilities, so we do not need to implement strong AGN feedback in this model. Our results are also in good agreement with the observed stellar mass function from $z$$\simeq$6.0 to $z$$\simeq$0.5.}{Our model offers the flexibility to test the impact of various physical processes on the regulation of star formation on a representative population of galaxies across cosmic times. Thermal instabilities and the cascade of turbulent energy in the dense gas phase introduce a delay between gas accretion and star formation, which keeps galaxy growth inefficient in the early Universe. The main results presented in this paper, such as stellar mass functions, are available in the GALAKSIENN library.}


\maketitle

%
%

\section{Introduction}

Galaxies are defined by their stellar populations --- the \lq when and where\rq\  their stars formed. Therefore, if models are to capture accurately the process of galaxy formation and evolution, researchers must determine how star formation is regulated locally and globally in galaxies. However, star formation is one of the most challenging processes to characterize in galaxy evolution models, essentially because the formation of stars involves many non-linear processes that occur over a large range of temporal and spatial scales in e.g., density, velocity, and magnetic field strength and regularity \citep{Kritsuk_2013, Krumholz_2005}. Observations show that star formation in galaxies is a very inefficient process, with typically 0.1\%-10\% of the available gas being converted into stars per local free-fall time \citep[e.g.,][]{Lada_2010, Hennebelle_2012, Agertz_2015, Lee_2016}. On the other hand, numerical simulations of molecular clouds indicate that the star-formation efficiency is highly dependent on how ionization and kinetic feedback is injected into the interstellar medium \citep[ISM; e.g.,][]{Krumholz_2012, Gatto_2015, Hennebelle_2014, Dale_2015, Geen_2017, Gavagnin_2017}. 

The difficulty in simulating feedback-regulated star formation, as well as the absence of a detailed physical description of processes responsible for large-scale feedback (whether driven by active galactic nuclei, AGN, or intense star formation or both) and gas accretion onto galaxies, make the global regulation of star formation one of the greatest challenges in modelling galaxy formation and evolution. Recently, there is growing observational and theoretical evidence that turbulent pressure injected by young stars is comparable to gravitational pressure in distant disks, enabling self-regulated star formation with low efficiency \citep{Lehnert_2013}. Warm diffuse and cold dense gas evolve under the influence of compression by passing spiral arms, thermal and gravitational instabilities, and supernovae-driven shocks. High resolution hydrodynamic simulations, with implementation of sub-grid turbulence models \citep{Schmidt_2013, Semenov_2016}, show complex, multi-phase, turbulent structures within the ISM with a realistic global Kennicutt-Schmidt relation on kpc scale, and gas depletion times in star forming regions over scales of ~10-50~pc consistent with observations. In these models, the global gas depletion time is long ($\tau_{depl} = M_g / \dot{M_{\star}} \approx 1 - 10$~Gyr, where $M_g$ and $\dot{M_{\star}}$ are respectively the gas mass and the star-formation rate),  because the gas spends most of the time in a state that does not rapidly lead to star formation. The gas is recycled many times, since the lifetime of star-forming clouds or the local gas depletion time in star forming regions is typically $1-500$~Myr. This wide range of local depletion times which lead to significant gas cycling and re-cycling is due to dynamical disruption, dispersal by feedback \citep{Raskutti_2016, Semenov_2017}, and supersonic turbulence \citep{Guillard_2009}. This complexity must be captured in some way in galaxy evolution models to generate realistic galaxies.

In high resolution hydrodynamical simulations \citep[e.g.,][]{Hopkins_2014, Schaye_2015, Kimm_2017, Mitchell_2018} or in semi-analytical models \citep[SAMs; e.g.,][]{Cole_1991, Cole_2000, Hatton_2003, Baugh_2006, Croton_2006, Cattaneo_2006, Somerville_2008, Guo_2011, Henriques_2013}, depending on the model and the mass of the galaxy, different prescriptions for various feedback mechanisms have been invoked to regulate star formation. For low stellar masses, $\Ms < 10^{9}\Msun$, feedback is used to either regulate gas accretion, mostly by heating the gas through photo-ionisation \citep[e.g.,][]{Doroshkevich_1967, Couchman_1986, Ikeuchi_1986, Rees_1986}, or ejecting the gas using the mechanical energy generated by supernovae \citep[SN, e.g.,][]{White_1978}. For high stellar masses, $\Ms > 10^{10}\Msun$, models rely on the action of Super-Massive Black-Holes (SMBH) to inhibit gas accretion onto galaxies. A significant fraction of the power generated by AGN is used to limit the cooling of the hot gas phase surrounding the galaxy.

Despite including some of these processes to regulate the gas content of galaxies, galaxy evolution models fail to reproduce the star-formation histories and physical properties of galaxies, mainly because a robust theory of star formation and dynamical coupling between gas phases are still lacking. As explained in \cite{Cousin_2015a}, current semi-analytical model overestimate the number of low-mass galaxies, especially at high redshift ($z > 2.0$), where the gap between models and observations is roughly an order of magnitude or more. For high mass galaxies, AGN feedback, initially used to limit the growth of massive galaxies at low redshift, has also an impact on the star-formation rate and history at higher redshift. Consequently, even the massive galaxies, those with $\Ms \simeq 10^{11}\Msun$ observed at redshifts greater than 3, are not robustly reproduced in current models.

We present here a new semi-analytical model \GAS\,--- the Galaxy Assembler from dark-matter Simulation --- which is based, in part, on previous version described in \cite{Cousin_2015b} and \cite{Cousin_2016}. This paper (paper I) provides an overview of the physical processes considered in \GAS\, and how they are implemented as phenomenological rate equations. In addition, we have two complementary companion papers: \GAS\,II: in which we describe and model the mechanisms that leads to dust attenuation of galaxian light and \GAS\,III in which we explore the panchromatic emission of galaxies from the FUV to the sub-millimetre bands. 

In this first paper we focus mainly on the regulation of star formation. In previous models, we have adopted an \textit{ad-hoc} recipe to generate a delay between accretion and star formation \citep{Cousin_2015a}. Here we implement a physical prescription based on the inertial cascade of turbulent energy from large to small scales. Accreted gas onto galaxies is initially considered as mainly diffuse. We compute the mass fraction of the gas subject to phase separation and fragmentation following \citet{Sharma_2012} and \citet{Cornuault_2018} and references therein, which also depends on the disk properties. Star formation occurs in the fragmented gas at a scale of 0.1~pc. In a large set of semi-analytical models \citep[e.g.,][]{Croton_2006, Cattaneo_2006, Somerville_2008}, star formation in massive galaxies is regulated by a strong reduction, or even complete suppression, of gas accretion by AGN feedback. In our new model, we do not need efficient AGN feedback, but instead our regulation process is a natural outcome of both the growth of thermal instabilities in the hot halo phase, and the dissipation of turbulent energy within the denser, fragmented gas reservoir. Those two processes delay gas accretion onto galactic disks and star formation.

The paper is organised as following. In Sect.~\ref{sec:accretion}, we provide a brief description of the dark-matter simulation we use, as well as the prescription we adopt to implement baryonic accretion rates. In Sect.~\ref{sec:the_turbulent_inertial_cascade}, we focus on the turbulent inertial cascade. We describe how we compute the energy and mass transfer rates between physical scales, we define and compute the gas fragmentation timescale, and show that it is a key parameter for the regulation of star formation in low-mass galaxies. In Sect.~\ref{sec:Gas_cycle}, we describe our model for the gas cycle as it accretes onto galactic disks, from diffuse accreted gas to potentially star forming gas. In Sect.~\ref{sec:feedbacks}, we describe our implementation of SN and AGN feedback. Sect.~\ref{sec:thermal_instabilities} focuses on the thermal instabilities arising in the hot gas phase in the halo. We assume that gas accretion onto galaxies is limited by turbulent mixing in the range of radii, where thermal instabilities develop because gas acquires large velocity dispersions. This allows us to define an effective cooling rate. In section \ref{results} we present and discuss our results, mainly focusing on the evolution of the galaxy stellar mass function with redshift, and the evolution of relevant timescales of physical processes (e.g., gas cooling, fragmentation, and orbital timescales). We also discuss the impact of our implementation of thermal instabilities on the quenching of massive galaxies in massive halos.

%
%

\section{From dark-matter to baryons}
\label{sec:accretion}

\subsection{Dark-matter}
\label{sec:dark_matter}

\GAS\, is built upon a set of dark-matter merger trees extracted from a pure N-body simulation. The current simulation uses WMAP-5yr cosmology \citep[$\Omega_m = 0.28$, $\Omega_{\Lambda} = 0.72$, $f_b = 0.16$, $h = 0.70$,][]{Komatsu_2009} with a volume of $[100/h]^3 Mpc$ in which $1024^3$ particles evolve. Each particle has a mass of $m_p = 1.025~10^8~\Msun$. Halos and sub-structures (satellites) are identified by using the \verb?HaloMaker? code \citep{Tweed_2009}. In our merger trees, we only consider halos with at least 20 dark-matter particles leading to a minimal dark-matter halo mass of $2.050\times10^9~\Msun$. Dark-matter halos grow from smooth accretion. The dark-matter accretion rate, $\dot{M}_{dm}$, only includes particles that are newly detected in the halo, and that have never been previously identified in another halo.

\subsection{Baryonic accretion}
\label{sec:gas_accretion}

\begin{figure}[t]
    \centering
        \centering
        \includegraphics[width=\linewidth]{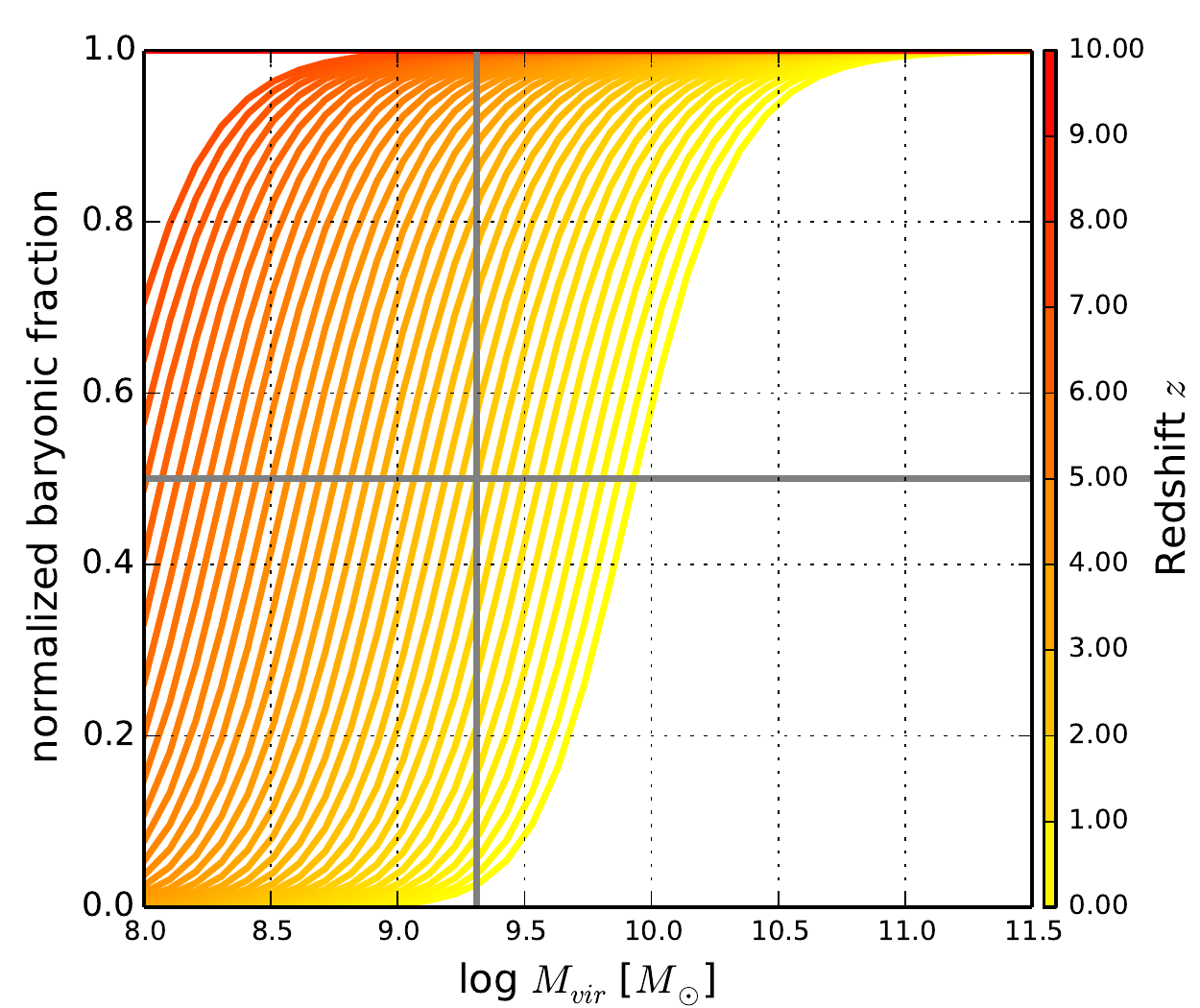}
    \caption{\ftns{The relative mass fraction of baryons which are accreted smoothly. The fraction is plotted as function of both the dark-matter virial mass and the redshift. The redshift of each curve is colour coded by the color bar on the right. Each curve, starting at $z = 0.0$, is separated by $\Delta z = 0.1$. For $z > z_{reion} = 7.0$, we set $f_b^{ph-ion}/f_b = 1.0$. The grey vertical bar marks the minimum halo mass in our dark-matter model. The grey horizontal line marks $f = 0.5$.}}
    \label{fig:normalized_baryonic_fraction}
\end{figure}

Based on the dark-matter accretion rate $\dot{M}_{dm}$, the accretion rate of baryons is defined by,
\begin{equation}
\dot{M}_b = f_b^{ph-ion}(M_h,z)\dot{M}_{dm} \ ,
\label{eq:gas_accretion}
\end{equation}
\noindent
where $f_b^{ph-ion}(M_h,z)$ is the effective baryonic fraction. The baryon accretion rate depends on the gas ionisation state and we adopt here the photo-ionisation model based on the \cite{Gnedin_2000} and \cite{Kravtsov_2004} prescription, but using the effective filtering mass given by \cite{Okamoto_2008}. We assume that re-ionisation occurs at $z = 7.0$, we therefore set $f_b^{ph-ion}(M_h,z>7) = f_b = 0.16$.

Fig. \ref{fig:normalized_baryonic_fraction} shows the normalised mass baryonic fraction that is associated with the dark matter smooth accretion (we assume a universal baryonic fraction $f_b=0.16$). Following the minimal dark-matter halo mass used in our model, the main impact ($f_b^{ph-ion}/f_b < 0.5$) of the photo-ionisation prescription occurs at low redshift ($z < 0.9$).

As initially proposed by \cite{Khochfar_2009}, in \GAS\, we define two different modes of accretion, a cold\footnote{The temperature of the accreted gas in this mode is close to $10^4$K.} mode, and a hot mode. Depending on the dark-matter halo mass, the fraction of accreted, hot gas is computed as
\begin{equation}
f_{sh}(M_{vir}) = \dfrac{1}{2}\left[1+\text{erf}\left(\text{log}M_{vir} - \text{log}M_{mix}\right)\right]\ , 
\label{eq:shock_heated_fraction}
\end{equation}

\noindent
where $M_{mix}$ is the transition mass when the cold and hot gas mass accretion rates are equal and $M_{vir}$ is the halo virial mass. The evolution of the hot gas fraction is inspired by the study of \cite[][see their Eqs. 24 and 25]{Lu_2011}, but we do not account for evolution with redshift since it is very weak and does not strongly impact how the gas is accreted in our model.

The baryonic accretion is divided in two parts that feed two different reservoirs: $M_{cold}$ for the cold mode and $M_{hot}$ for the hot mode. Both the cold and the hot reservoirs are fed by metal-free gas. During the evolution of any galaxy, metal-rich ejecta coming from the galaxy are added to the hot reservoir. The metal content of the hot gas therefore depends directly on the rates and timescales over which galaxies create metals. The metallicity of the hot reservoir evolves with time. The chemo-dynamical model included in \GAS\, tracks the abundance of the main elements in the gas phase. The production and the re-injection of these metals are taken into account for stars with initial masses between 0.1 M$_{\sun}$ and 100 M$_{\sun}$ over metallicities from zero to super-solar.

\subsubsection{The cold accretion mode}

For the cold accretion mode, we assume that gas falls directly onto the galaxy \citep{Dekel_2006}, and we compute the cold accretion rate via:
 \begin{equation}
    \dot{M}_{streams} = \dfrac{M_{cold}}{2t_{dyn}} \ ,
    \label{eq:ff_rate}
\end{equation}
where $M_{cold}$ is the mass stored in the cold reservoir and $t_{dyn}$ is the dynamical time of the dark-matter halo: $t_{dyn} = r_{vir} / v_{vir}$. 

\begin{figure}[t!]
    \centering
        \centering
        \includegraphics[width=\linewidth]{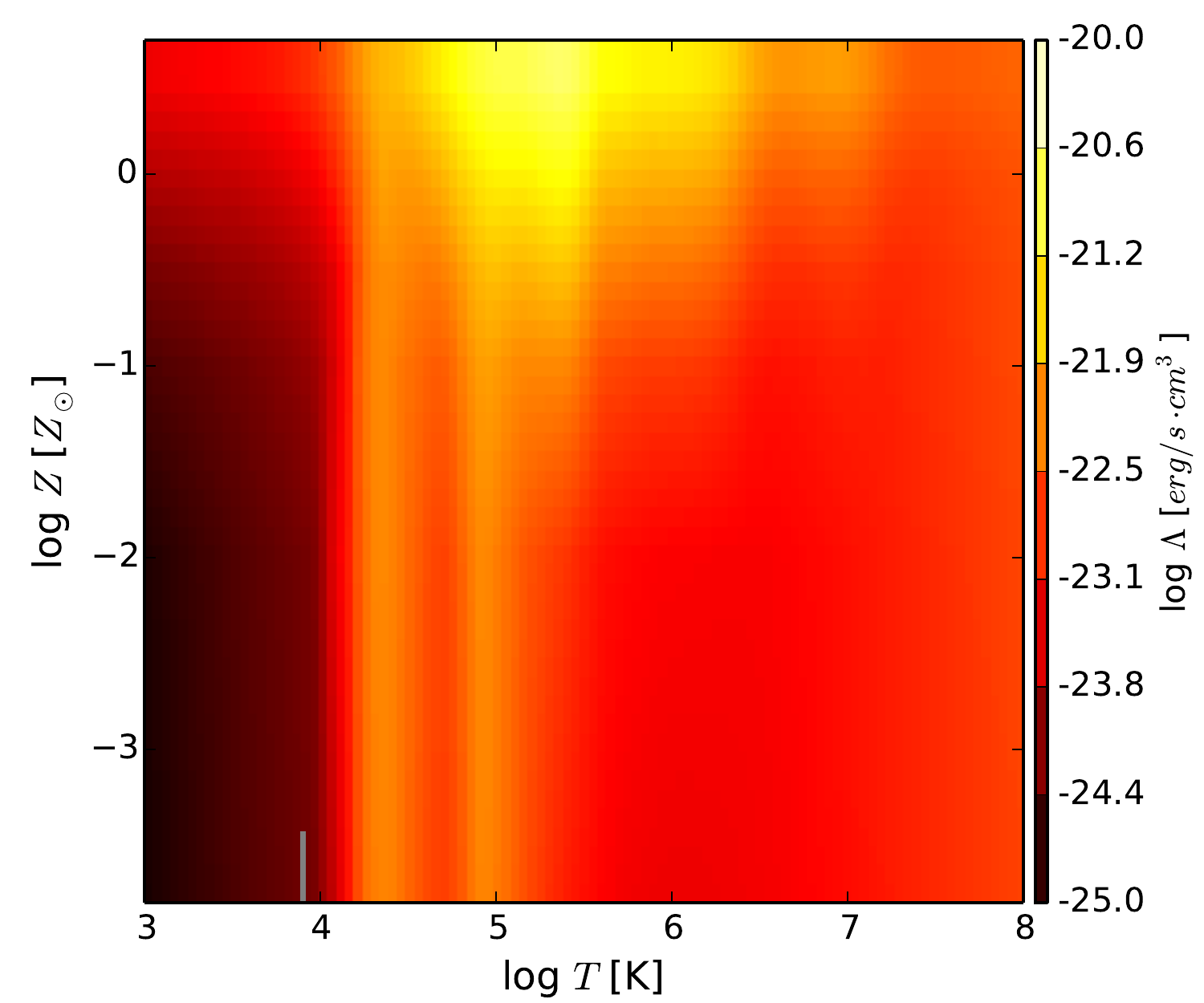}
    \caption{Cooling efficiency, $\Lambda(T_{hot},Z_{hot})$, as a function of both gas temperature and gas metallicity \citep[see][]{De_Rijcke_2013}. The values of $\Lambda(T_{hot},Z_{hot})$ are color coded using the color bar on the right (in units of erg s$^{-1}$ cm$^{-3}$).}
    \label{fig:thermal_cooling_eff}
\end{figure}

\subsubsection{Hot accretion: Radiative cooling}
\label{sec:hot_mode}

As in the previous versions of this model \citep{Cousin_2015b, Cousin_2016}, we assume that the hot gas surrounding the galaxy is confined in the potential well of the dark-matter halo and in hydrostatic equilibrium. The hot gas density profile $\rho_{hot}(r)$ is computed following the prescriptions in \cite{Suto_1998}, \cite{Makino_1998}, \cite{Komatsu_2001} and \cite{Capelo_2012}. We assume for such hot atmosphere a constant temperature $T_{hot}$, and a average gas metallicity $Z_{hot}$. For more details, please refer to \citet[][Sect.~4]{Cousin_2015b}. Radiative cooling and the associated gas condensation is computed using the prescription in \cite{White_1991}. The cooling time of the hot gas phase is defined -- as a function of the radius -- as,
\begin{equation}
   t_{cool}(r) = 10.75\dfrac{\mu m_p T_{hot}}{\rho_{hot}(r)\Lambda[T_{hot},Z_{hot}]}\, .
   \label{eq:cooling_time_function}
\end{equation}

In previous versions of this model, we adopted the \cite{Sutherland_1993} cooling efficiencies $\Lambda(T,Z)$. In the present version, we use those computed by \cite{De_Rijcke_2013}, tabulated as a function of gas temperature and metallicity, which we interpolate between $T = 10^3$K and $10^8$K, and for gas metallicities over the range 10$^{-4}Z_{\odot}$ and $2Z_{\odot}$. We assume the solar metal mass fraction of $Z_{\odot} = 0.02$. Fig.~\ref{fig:thermal_cooling_eff} shows the cooling efficiency as a function of both gas temperature and average gas metallicity.

At a given time, the mass of warm gas that can condensate and feed the galaxy is enclosed within the cooling radius, $r_{cool}$. This radius is calculated using the cooling time equation:
\begin{equation}
	t(r_{cool}) = \mathcal{T}_{cool}^{hot}
    \label{eq:cooling_radius}
\end{equation}

In this equation, $\mathcal{T}_{cool}^{hot}$ is a ``cooling clock'', which is a measure of the effective cooling time of the hot gas. After each time-step $\Delta t$, the mass-weighted cooling time is updated via:
\begin{equation}
   \mathcal{T}_{cool}^{hot,n} = \underbrace{\left(\mathcal{T}_{cool}^{hot,n-1} + \Delta t\right)\left(1-\frac{\Delta M}{M}\right)}_{\text{hot halo gas}} + \underbrace{\frac{\Delta t}{2}\frac{\Delta M}{M}}_{\text{newly incoming hot halo gas}} \ ,
   \label{eq:cooling_clock}
\end{equation}
\noindent
where $M$ is the total mass in the hot phase after the latest time-step, $\Delta t$. $\Delta M$ is the net mass variation of hot gas during this time-step (accretion - ejection). We assume that cooling takes place during the overall last time-step for the gas already in the hot atmosphere. However, accretion is continuous\footnote{Using the adaptive time-step method described in \citep{Cousin_2015b}, we ensure that all exchange rates between reservoirs are constant during each time-step.}, and incoming gas also starts to cool. Therefore, taking into account the time for the incoming gas to enter the hot phase, this new gas cools during only half a time-step, on average. Therefore, the effective cooling time of the hot gas halo can increase or decrease between two time-steps depending on the relative fraction of halo gas to incoming gas. During mergers, the cooling clock of the remnant hot phase is set to the value of the most massive progenitor at the time of the merger.

Knowing the cooling radius $r_{cool}$ (deduced from Eq.\ref{eq:cooling_radius}), we can write the condensation rate of the gas:
\begin{equation}
  \ds \dot{M}_{cool} = \dfrac{v(r_{cool})}{2r_{cool}}\int_0^{r_{cool}}\rho_{hot}(r)r^2dr.
  \label{eq:cooling_rate}
\end{equation}

\noindent
The mass within $r_{cool}$ decreases with a timescale $r_{cool}/v(r_{cool})$, where $v(r_{cool})$ is the circular velocity of the dark-matter halo measured at $r = r_{cool}$. We assume that the hot atmosphere extends up to the virial radius $r_{vir}$ of the dark-matter halo, thus the cooling radius cannot be larger than $r_{vir}$.

\begin{figure}[t]
  \begin{center}
    \includegraphics[width=\linewidth]{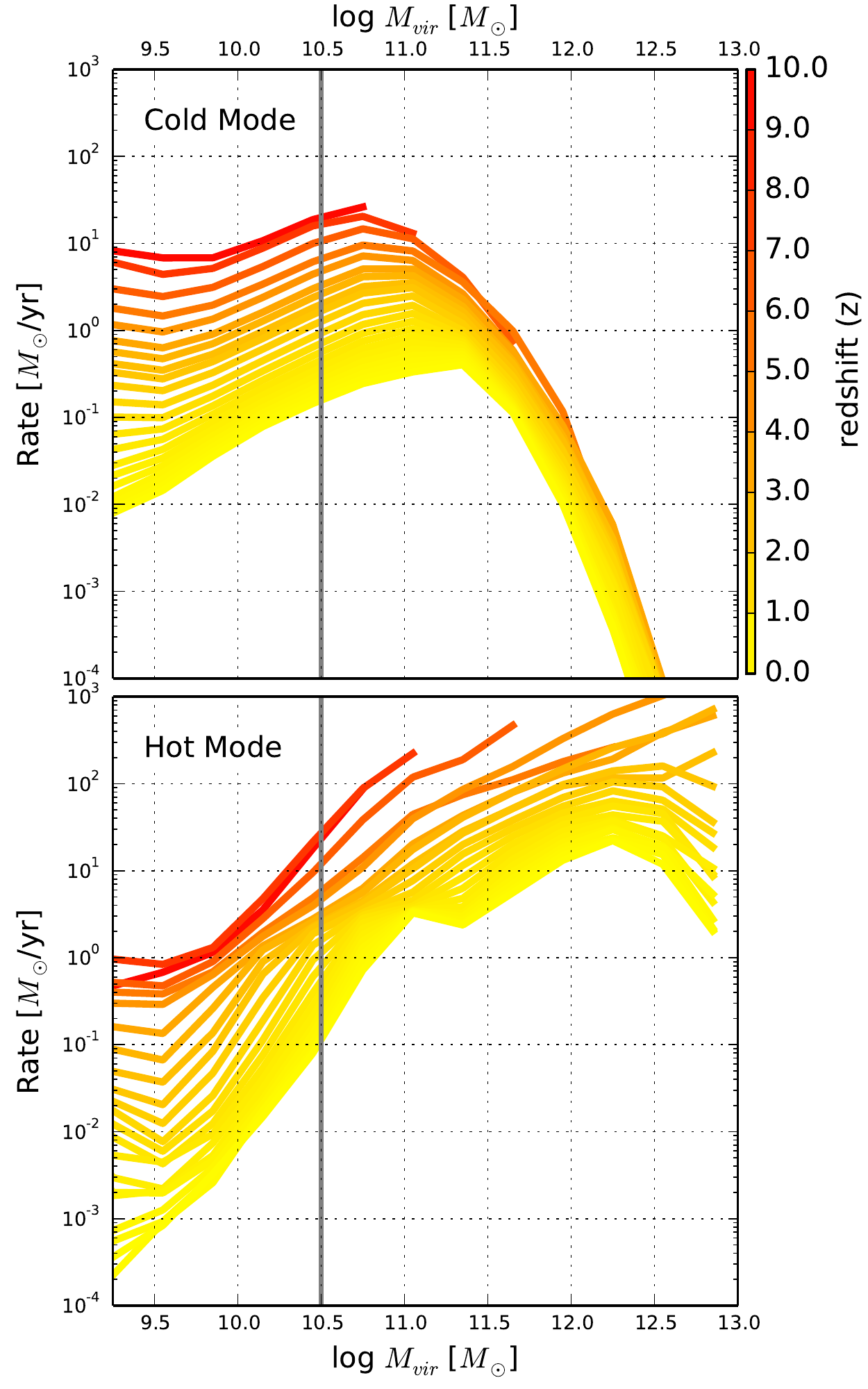}
  \caption{\ftns{Mean accretion rates into the galaxy halo galaxy from cold (upper panel) and the hot mode accretion (lower panel) as a function of virial mass. The mean values are computed by selecting only halos which are actively accreting gas. Overall, as expected, the both the cold and hot mode accretion rates decline with decreasing redshift but the relative fraction of hot mode accretion increases with decreasing redshift and for the most massive halos ($M_{h}>10^{10.5}\Msun$). This trend is the result of the cooling regulation related to thermal instabilities. Colour code indicates redshift. The grey solid vertical line marks the dark-matter halo mass where contributions of cold and hot modes are equivalent.}}
  \label{fig:acc_rate}
  \end{center}
\end{figure}

Fig. \ref{fig:acc_rate} shows the average galaxy gas accretion rates produced by the two different modes as a function of both the dark-matter virial mass and the redshift. At low dark-matter halo masses, $M_{vir} < 10^{10.5}\Msun$, accretion on the galaxy is dominated by the cold mode. The contribution to the accreted mass of the hot mode increases progressively with halo mass. Around $M_{vir} \sim 10^{10.5}\Msun$, the contribution of the two modes are equal. This transition occurs at approximately the same halo mass for all redshifts we considered. For both the cold mode and the hot mode accretion, the average accretion rate decreases with the redshift. For dark-matter halos with $M_{h} = 10^{10.5}\Msun$, the average accretion due to the cold mode decreases from 20$\Msun$ yr$^{-1}$ at $z \simeq 9.0$ to 0.2$\Msun$ yr$^{-1}$ at $z = 0.3$; for the hot mode, accretion rate decreases from 30$\Msun$ yr$^{-1}$ to less than 0.3$\Msun$ yr$^{-1}$ between $z\simeq 9.0$ and $z = 0.3$. 

\begin{table*}[th!]
  \begin{center}
      \begin{tabular}{lllll}
        \hline
        \textbf{symbol} & \textbf{definition} & \textbf{Eq/Sect} & \textbf{values} \\
        \hline
        \\
        & -- Reference Masses --\\
        $M_{h,min}$          & Minimal dark-matter halo mass (20 particles) & Sect. \ref{sec:dark_matter} & $2.05\times 10^{9}\Msun$ \\
        $M_{mix}$            & Transition mass from cold to hot accretion mode & Sect. \ref{sec:gas_accretion}, Eq. \ref{eq:shock_heated_fraction} & $10^{11}\Msun$\\
        $M_{\bullet}^{init}$ & Initial super massive black hole mass (seed) & Sect. & 300 $\Msun$ & \\
        \\
        & -- Thermal instability --\\
        $\varepsilon_{TI}$ & Thermal instability propagation efficiency & Sect. \ref{sec:Mixing_zone}, Eq. \ref{eq:TI_radius} & 0.63\\
        \\
		& -- Feedback: Repartition --\\
		$f_{ME}$ & Fraction of SMBH infall rate converted in power & Sect. \ref{sec:AGN_feedback}, Eqs. \ref{eq:AGN_ejecta_rate} & 0.1\\
		$f_{k,SN}$ & Kinetic fraction of SN energy & Sect. \ref{SN_ejecta_rate_EQ}, Eq. \ref{SN_ejecta_rate_EQ} & 2/3 & \\
		$f_{w}$ & Fraction affected to large scale wind & Sect. \ref{SN_ejecta_rate_EQ}, Eq. \ref{SN_ejecta_rate_EQ} & 0.2 &\\
		$f_{Th,SN}$ & Thermal fraction of non kinetic SN energy & Sect. \ref{sec:thermal_luminosity_power}, Eq. \ref{eq:SN_disruption_rate} & 1/2 &\\
		$f_{k,AGN}$ & Kinetic fraction of AGN energy & Sect. \ref{sec:Large_scale_ejecta}, Eq. \ref{eq:Vjet} & $10^{-3}$\\
		$f_{Th,AGN}$ & Thermal fraction of non kinetic AGN energy & Sect. \ref{sec:thermal_luminosity_power} & 0.6\\
		 \\
        & -- Accretion --\\
        $f_{frag}^{in}$ & Fraction (in mass) of gas which is already fragmented when accreted & Sect. \ref{sec:Gas_cycle} & 1/3 \\
        \\
        & -- Turbulent kinetic energy budget  --\\
        $f_{incr}$ & Fraction of the rotational energy associated with the newly & Sect. \ref{sec:average_velocity_dispersion}, Eq. \ref{eq:Updated_kinetec_energy_budget} & 1/3 \\
                   & diffuse accreted gas converted in turbulent kinetic energy & & \\
        $f_{disp}$ & Fraction of the turbulent kinetic energy budget & Sect. \ref{sec:average_velocity_dispersion}, Eq. \ref{eq:Updated_kinetec_energy_budget} & 1/2 \\
        		   & dissipated per dynamical time step & & \\
        \\		
		& -- Feedback: Additional parameters --\\
		$v_{w}$ & Large scale wind velocity & Sect. \ref{sec:Large_scale_ejecta}, Eq. \ref{SN_ejecta_rate_EQ} & [100,200] km/s\\ 
		$E_{SN}$   & Total SN energy & Sect. \ref{sec:SN_feedback} & $10^{44}$J  & \\
		$\eta_{\bullet}$ & Inflow-outflow ratio for SMBH activity & Sect.~\ref{sec:Large_scale_ejecta} & 1.0 & \\ 
		\\
		$\eta_{m}$ & Minimal mass ratio for major merger events & & 1/3 \\	
        \hline
      \end{tabular}
  \end{center}  
  \caption{\footnotesize{Definitions, values and associated references of parameters used in the current model.}}
  \label{full_model_parameters_TAB}
\end{table*}

%
%

\section{The turbulent inertial cascade}
\label{sec:the_turbulent_inertial_cascade}

\begin{table*}[th]
  \begin{center}
      \begin{tabular}{llll}
        \hline
        \textbf{symbol} & \textbf{definition}  & \textbf{values} & \textbf{Refs}\\
        \hline
        $k_{\star}$      & Star-formation wave number  & $  1/l_{\star} = 10~pc^{-1}$ & \cite{Andre_2013a} \\ 
        $\mu_{\star}$    & Mass surface density threshold for star-formation   & $ 150~\Msun/pc^2$ & \cite{Lada_2012}\\
        $\sigma_{\star}$ & Velocity dispersion at the star-formation scale  & $ 0.3~km/s$ & \cite{Arzoumanian_2013} \\
        $a$              & Slope index of the Larson surface density law & $  1/5$ & \cite{Romeo_2010}\\ 
        $b$              & Slope index of the Larson 1D velocity dispersion law & $  3/5$ & \cite{Romeo_2010}\\
        \hline
      \end{tabular}
  \end{center}  
  \caption{\footnotesize{Definitions, values and associated references of the parameters used in the turbulent cascade scaling relations (Sect.~\ref{sec:the_turbulent_inertial_cascade} and Eq. \ref{eq:Larson_mu_sig}.)}}
  \label{tab:turbu_parameter_list}
\end{table*}

Both observations and numerical simulations of the ISM at high spatial resolutions show that turbulence plays a fundamental role in star formation \citep[e.g.,][]{Bergin_2007, Miville-Deschenes_2010, Renaud_2014}. Turbulence controls the rate at which kinetic energy is dissipated \citep[e.g.,][]{Krumholz_2005,Padoan_2011,Federrath_Klessen_2012} and leads to development of multi-phase morphological structures in the gas \citep[e.g.,][]{Andre_2013a, Levrier_2018}. Recent numerical models of galaxy formation have adopted a gravo-turbulent sub-grid model for star formation \citep[e.g.,][]{Hopkins_2012, Kimm_2017}, but those time-consuming simulations are limited to small cosmological volumes. In this section, we develop our modelling of the mass and energy transport from the large scale of injection to the small dissipative scale, through a hierarchy of structures, using a phenomenological prescription for the cascade of turbulent energy. This allows us to compute the mass of gas that may form stars and to test the impact of these phenomenon on the properties of galaxies. In our prescription, the mass of gas that is able to form stars is the mass of gas reaching the dissipation scale $l_{\star}$ (see Sect.~\ref{sec:star-formation_threshold}), calculated from the mass flow rate between scales, under the assumption of a constant energy transfer (Sect.~\ref{sec:mass_transfer_rate}). 

\subsection{Self-similar scaling relations and energy transfer rate}
\label{sec:star-formation_threshold}

We assume that the two \citet{Larson_1981} self-similar scaling relations are satisfied to compute the mass and energy transfer rates during the inertial turbulent cascade. They link the mass surface density $\mu$ and the 1D-velocity dispersion $\sigma$, respectively, to the wave number $k=1/l$:
\begin{equation}
	\mu_k = \mu_{\star}\left(\dfrac{k}{k_{\star}}\right)^{-a} \ \rm{and} \ \  \sigma_k = \sigma_{\star}\left(\dfrac{k}{k_{\star}}\right)^{-b} \ ,
    \label{eq:Larson_mu_sig}
\end{equation} 
where $k_{\star}$ is the wave number associated with the star-formation scale, $l_{\star} = 1/k_{\star}$. $\mu_{\star}$ and $\sigma_{\star}$ are two normalisation parameters and $a$ and $b$ are the two slope indices; both depend on the gas phase considered \citep{Fleck_1996, Hoffmann_2012}. The second self-similar scaling relation gives the energy transfer rate per unit volume, as in \citet{Kritsuk_2013}: 
\begin{equation}
  \dot{e} \propto \rho(k)\sigma(k)^3k \propto \mu(k)\sigma(k)^3k^2~[M\cdot L^{-1} \cdot T^{-3}] \ .
  \label{eq:constant_transfer_rate_1}
\end{equation}   

We set the energy transfer rate $\dot{e}$ to be constant \citep[e.g.,][]{Hennebelle_2012}, which gives a relation between the two slopes of Larson's relations: $b = \frac{1}{3}(2-a)$. Following \citet{Romeo_2010}, we assume $(a,b)=(1/5,3/5)$. These values reproduce observations of the dynamics of clouds at the atomic-molecular transition. The energy transfer rate during the inertial cascade is now proportional to:
\begin{equation}
  \dot{e} \propto \mu_{\star}\sigma_{\star}^3 k_{\star}^{2} \ .
  \label{eq:constant_transfer_rate_2}
\end{equation}  

The normalisation parameters used in Eq.~\ref{eq:Larson_mu_sig} are listed in Table~\ref{tab:turbu_parameter_list}, and correspond to standard values for dense, supersonic, compressible gases \citep{Jog_1984a,Jog_1984b,Romeo_2010,Hoffmann_2012}. We assume $l_{\star} = 0.1~pc$ for the star-formation scale, which corresponds to the characteristic width of interstellar filaments hosting pre-stellar cores \citep[e.g.,][]{Andre_2013a, Palmeirim_2013}. Although the detailed physical interpretation of this width is still debated, this scale is of the order of the scale below which the turbulence becomes subsonic in star-forming filaments \citep[e.g.,][]{Padoan_2001}. At this scale, we adopt a typical velocity dispersion $\sigma_{\star} = 0.3$~km s$^{-1}$ \citep[e.g.,][]{Orkisz_2017}, a value slightly higher than the speed of sound for a molecular gas at 10~K ($c_s(10 K) = 0.2$ km s$^{-1}$), which corresponds to the observed transition between bound and unbound filaments \citep{Arzoumanian_2013}. The critical mass surface density above which the gas is gravitationally unstable and converted into stars is set to $\mu_{\star} \simeq 150~\Msun$ pc$^{-2}$ \citep{Gao_Solomon_2004, Heiderman_2010, Lada_2012}. 

\begin{figure}[t!]
  \begin{center}
    \includegraphics[width=\linewidth]{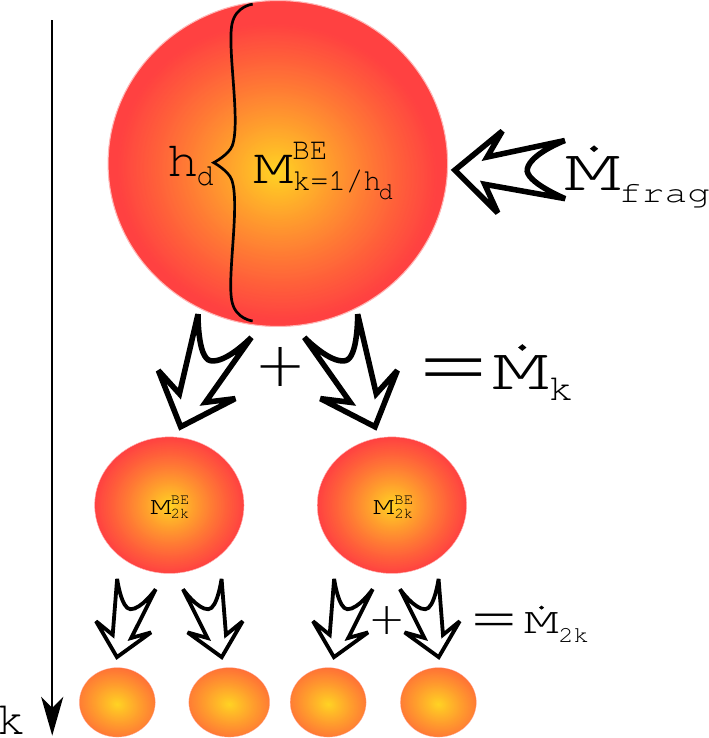}
  \caption{\ftns{Diagram illustrating how the gas fragments during the turbulent inertial cascade. When a structure which formed at the scale $k$ reaches the Bonnor-Ebert mass, it breaks into smaller structures at the next level, $2k$. The mass transfer rate, $\dot{M}_k$, between scales is given by Eq.~\ref{eq:transfer_rate}. The largest structures are fed at a rate, $\dot{M}_{frag}$, as discussed in Sect.~\ref{sec:diffuse_gas_reservoir} and given by Eq. \ref{eq:fragmenting_rate}.}}
  \label{fig:inertial_cascade}
  \end{center}
\end{figure}

\begin{figure}[t!]
\begin{center}
\includegraphics[width=\linewidth]{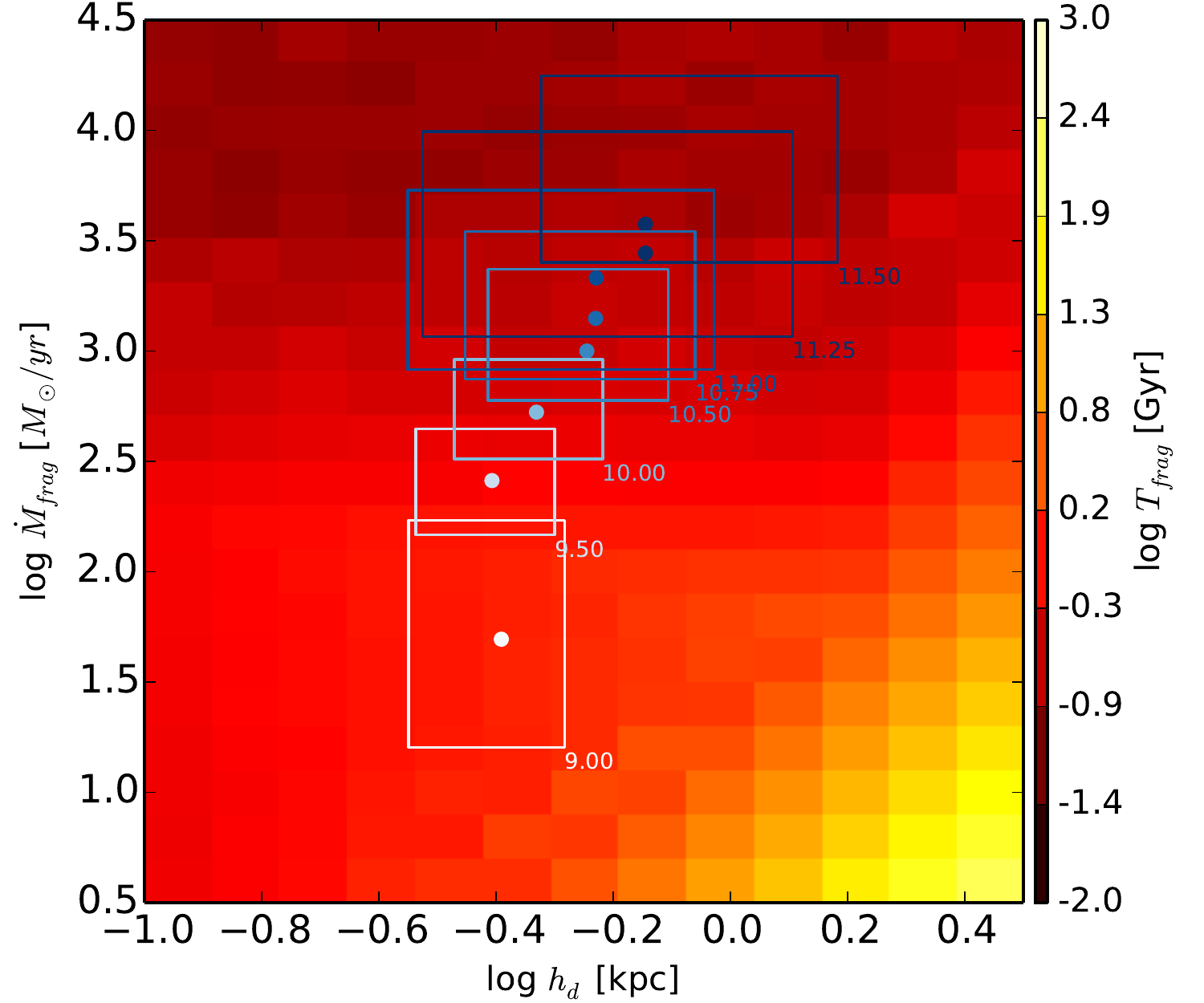}
\caption{\ftns{The gas fragmentation rate as a function of both the maximum energy injection scale, $l_{max}$, which we assume is equal to the disk scale-height, and the growth rate of GMCs (Eq. \ref{eq:fragmenting_rate}). Each coloured box shown in the diagram represents the gas fragmentation rate for a specific stellar mass bin as indicated by the labels at the bottom right corner of each box. The mass bins  range from $M_{\star} = 10^{9.0}M_{\odot}$ to $10^{11.5}M_{\odot}$. The limits of each box encompass the 15\% and 85\% percentiles of the distribution for each mass bin. Coloured points within each box indicated the value of the median of the distribution. The regions shown are for modelled galaxies at $z = 2.1$.}}
  \label{fig:fragmenting_timescale}
  \end{center}
\end{figure}

\subsection{Mass transfer rate during the inertial cascade}
\label{sec:mass_transfer_rate}

In star-forming galactic disks, the structure of the gas is observed to be self-similar over a wide range of scales, from a few kpc to the length scale $l_{\star}$ over which star formation occurs \citep{Dickman_1990, Miville-Deschenes_2010}. In this section, we compute the mass transfer rate from the energy transfer rate (Eq.~\ref{eq:constant_transfer_rate_2}). Starting from the scale of the disk scale-height, $h_d$, which is the largest possible injection scale, the gas mass is progressively distributed over smaller scales. Following the inertial cascade, when going from a scale $1/k$ to $1/2k$, large structures break into smaller ones. A diagram of this progressive fragmentation of the gas in the ISM of our modelled galaxies is shown in Fig.~\ref{fig:inertial_cascade}.

The Bonnor-Ebert mass at a given scale $1/k$, $M_k^{BE}$, sets the critical mass above which the structure becomes unstable and collapses into smaller structures \citep{Bonnor_1956}, until we reach the dissipation scale $l_{\star}$:
\begin{equation}
M_k^{BE} = 1.5\dfrac{\sigma^4_k}{G^2\mu_k}=1.5\dfrac{\sigma_{\star}^4}{G^2\mu_{\star}}\left(\dfrac{k}{k_{\star}}\right)^{-11/5}
\end{equation}

We derive the mass transfer rate $\dot{M}_k$ between wave numbers $k$ and $2k$, using the conservation of energy:
\begin{equation}
    \frac{3}{2}\dot{M}_k\sigma^2_k = \dot{e}V_k\text{E}\left(\dfrac{M_k}{M_k^{BE}}\right) \ , 
    \label{eq:transfer_rate}
\end{equation}
where $V_k =\frac{\pi}{6}k^{-3}$ is the volume of the cloud at scale $k$, $M^k$ is the total mass stored at scale $k$, and $\dot{e}$ is the constant kinetic turbulent energy transfer rate per unit volume (Eq.~\ref{eq:constant_transfer_rate_2}).

\subsection{Gas fragmentation timescale}
\label{sec:struturing_time scale}

We define the gas fragmentation timescale, $T_{frag}$, as the time needed to transfer all the gas from the disk scale-height, $h_d$, to the star-formation length scale, $l_{\star}$, assuming that the fragmented gas reservoir is fed at a constant rate $\dot{M}_{frag}$. During the process of fragmentation, we assume that the disk scale-height is constant. In practice, we track the mass of the initial Bonnor-Ebert sphere along the cascade given by Eq.~\ref{eq:transfer_rate}, until a steady-state is reached. 

Fig.~\ref{fig:fragmenting_timescale} shows our estimated gas fragmentation timescale $T_{frag}[h_d,\dot{M}_{frag}]$ as a function of the instantaneous disk scale-height $h_d$ and mass flow rate at which the largest structure is fed, $\dot{M}_{frag}$ (see Sect.~\ref{sec:diffuse_gas_reservoir} and Eq.~\ref{eq:fragmenting_rate}). Both the average disk scale-height and the GMC growth rate increase with the stellar mass, and the average fragmentation timescale decreases. Obviously, gas accretion history, star formation, gas ejection modify those two parameters, $h_d$ and $\dot{M}_{frag}$. To take this into account, we will define an effective disk fragmentation timescale $\mathcal{T}_{frag}$, which depends on the history of the disk (see Sect. \ref{sec:fragmented_gas_reservoir} and Eq. \ref{eq:t_str}). 
 
%
%

\section{\GAS\, cycle: evolution of the gas reservoirs}
\label{sec:Gas_cycle}

\begin{figure}[t!]
\begin{center}
\includegraphics[width=\linewidth]{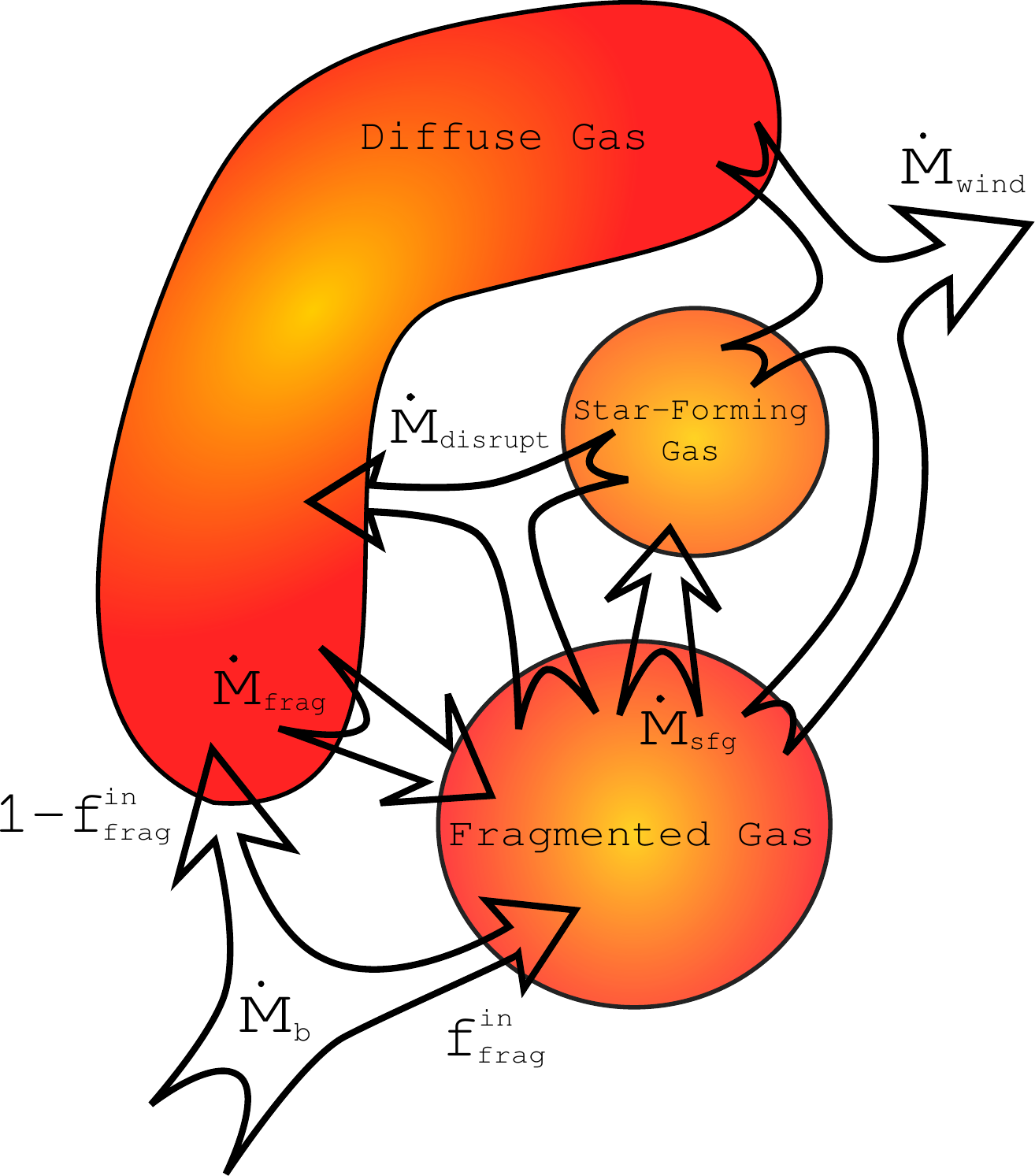}
\caption{Diagram illustrating the flow of mass between the three gas reservoirs we considered in our multi-phase \GAS\ model. The star-forming reservoir contains the mass that is immediately available for star formation. It is fed by a turbulent cascade and gas fragmentation at a mass flow rate, $\dot{M}_{sfg}$ (Eq.~\ref{eq:str_rate}). The gas reservoir fragments at a rate, $\dot{M}_{frag}$ (Eq.~\ref{eq:fragmenting_rate}). The diffuse gas reservoir is fed through the rates of two mechanisms, \textit{(1)} gas accretion rate, $(1-f^{in}_{frag}) \dot{M}_{b}$ (Eq.~\ref{eq:gas_accretion}), and, \textit{(2)} the rate at which the fragmented and star-forming gas is disrupted via the energy injected by SN and AGN, $\dot{M}_{disrupt}$ (Eqs.~\ref{eq:SN_disruption_rate} and \ref{eq:AGN_disruption_rate}). Each of these three gas reservoirs contributes to the outflow rate, $\dot{M}_{wind}$ (Eqs.~\ref{SN_ejecta_rate_EQ} and \ref{eq:AGN_ejecta_rate}).
} 
\label{fig:disc_structure}
\end{center}
\end{figure}

To model the gas cycle in galactic disks, we follow the mass content of three gas reservoirs: \textit{(i)} a diffuse gas reservoir; \textit{(ii)} a fragmented gas reservoir; and \textit{(iii)} a star-forming gas reservoir. In the following, we describe each of these gas reservoirs and their mass flow rates, of which we give a schematic view in  Fig.~\ref{fig:disc_structure}. We consider that the accreted gas onto the galaxy is multi-phase and we set the mass fraction of fragmented gas to $f_{frag}^{in} = 1/3$ \citep{vandeVoort2012}. We further discuss the physical origin and consequences of that assumption in Sect.~\ref{sec:discussion}.

\subsection{The diffuse gas reservoir}
\label{sec:diffuse_gas_reservoir}

The first main reservoir stores a diffuse ($\simeq$1 cm$^3$) and warm (10$^4$~K) gas which is traced by emission lines of the warm, ionised medium such as [O{\sc iii}], [N{\sc ii}], etc. \citep[e.g.,][]{Bethermin_2016, Laporte_2017, Suzuki_2017}. As illustrated in Fig.~\ref{fig:disc_structure}, the diffuse gas reservoir is fed by two different sources: \textit{(i)} the accretion of warm gas coming from both the cold and the hot mode, and \textit{(ii)} the disruption of fragmented gas by the injection of energy due to supernovae and/or an actively accreting super-massive black hole (see Sect.~\ref{sec:Disruption_rate}). 

The initial temperature of the warm, diffuse gas is assumed to be $10^4$K, and we compute its isobaric cooling using the same approach as with the hot gas phase (Sect~\ref{sec:hot_mode}). The effective cooling time of this phase $\mathcal{T}_{cool}^{unstr,n}$ is computed after each time step $\Delta t$, as the sum of the mass-weighted cooling times of the halo and the newly incoming gas:
\begin{equation}
   \mathcal{T}_{cool}^{diff,n} = \underbrace{\left(\mathcal{T}_{cool}^{diff,n-1} + \Delta t\right)\left(1-\frac{\Delta M}{M}\right)}_{\text{warm gas}} + \underbrace{\frac{\Delta t}{2}\frac{\Delta M}{M}}_{\text{newly incoming gas}} \ ,
   \label{eq:cooling_clock_diffuse_gas}
\end{equation}
where $M$ is the total mass of diffuse gas after the time-step $\Delta t$. $\Delta M$ is the mass increase of diffuse gas, coming from both accretion and disrupted gas coming from the fragmented phase. We assume that radiative cooling acts on the warm diffuse gas during all the previous time-step (Eq.~\ref{eq:cooling_clock_diffuse_gas}). However, for the newly incoming gas, we assume that the radiative cooling occurs only during half of the previous time-step (as for the hot gas phase). Therefore, the fraction of halo and freshly acquired gas can increase or decrease after each step.

The mass transfer rate between the diffuse gas phase to the fragmented gas phase $\dot{M}_{frag}$ is computed using the cooling timescale,

\begin{equation}
	\dot{M}_{frag} = (1-f_{frag})f_{Q}\phi_m\dfrac{M_{diff}}{t_{cool}^{warm}} \ ,
	\label{eq:fragmenting_rate}
\end{equation}

\noindent
where $M_{diff}$ is the mass of diffuse gas. The cooling timescale is computed as a function of the average metallicity $Z_{diff}$ (Eq.~\ref{eq:cooling_time_function}). We assume a temperature of $10^4$K and an average volume of the warm, diffuse gas component\footnote{The diffuse gas is assumed to evolve in a thick disk with scale-height $h_d$ and a total radius of the stellar disk equal to 11$r_d$. We assume that the warm diffuse gas can extend to a radius that is up to two times larger than total radius of the stellar disk.} $\mathcal{V}_{diff} = 22\pi r_d^2\,h_d$. The other parameters in Eq.~\ref{eq:fragmenting_rate} are computed as follows:

\begin{itemize}
\item $\phi_m$, the mass fraction of diffuse gas that condenses in an effective cooling time $\mathcal{T}_{cool}^{diff}$, depends on the characteristic cooling timescale of this reservoir $t_{cool}^{warm}$. $\phi_m$ was calculated in \citet{Cornuault_2018}, and for computational purposes we fitted their computation by an error function given by,

\begin{equation}
   \phi_m(x) = \dfrac{1}{2}\left[1+\text{ERF}\left(\dfrac{\log_{10}x-\log_{10}x_t}{\sqrt{s}}\right)\right] \ ,
   \label{eq:mass_filling_factor}
\end{equation}

\noindent
where $x=\mathcal{T}_{cool}^{diff} / t_{cool}^{warm}$. The best fit gives $x_t = 0.55$ and $s = 0.13$.

\item{$f_{frag}$, the fraction of the gas that is fragmented, is defined as,

\begin{equation}
f_{frag}=\frac{M_{frag}+M_{sfg}}{M_{diff}+M_{frag}+M_{sfg}},
\label{eq:fragmented_gas_fraction}
\end{equation}

\noindent
where $M_{diff}$, $M_{frag}$ and $M_{sfg}$ are the gas masses stored in the three different reservoirs (Fig.~\ref{fig:disc_structure}). This factor accounts for the fact that the more the gas is fragmented, the lower the mass flow to bound structures.}

\item{$f_{Q}$, the Toomre disk instability criterion, is calculated as,

\begin{equation}
f_{Q} = \text{MAX}\left[1.0~;\frac{Q_{crit}}{Q}\right] \ .
\label{eq:Toomre_factor}
\end{equation}	

This factor accounts for the fact that the mass flow rate to bound structures increases as the diffuse gas becomes gravitationally unstable. We adopt a standard value of $Q_{crit}$ = 1.0}.

\end{itemize}

\begin{figure*}[t!]
  \begin{center}
  \includegraphics[width=\linewidth]{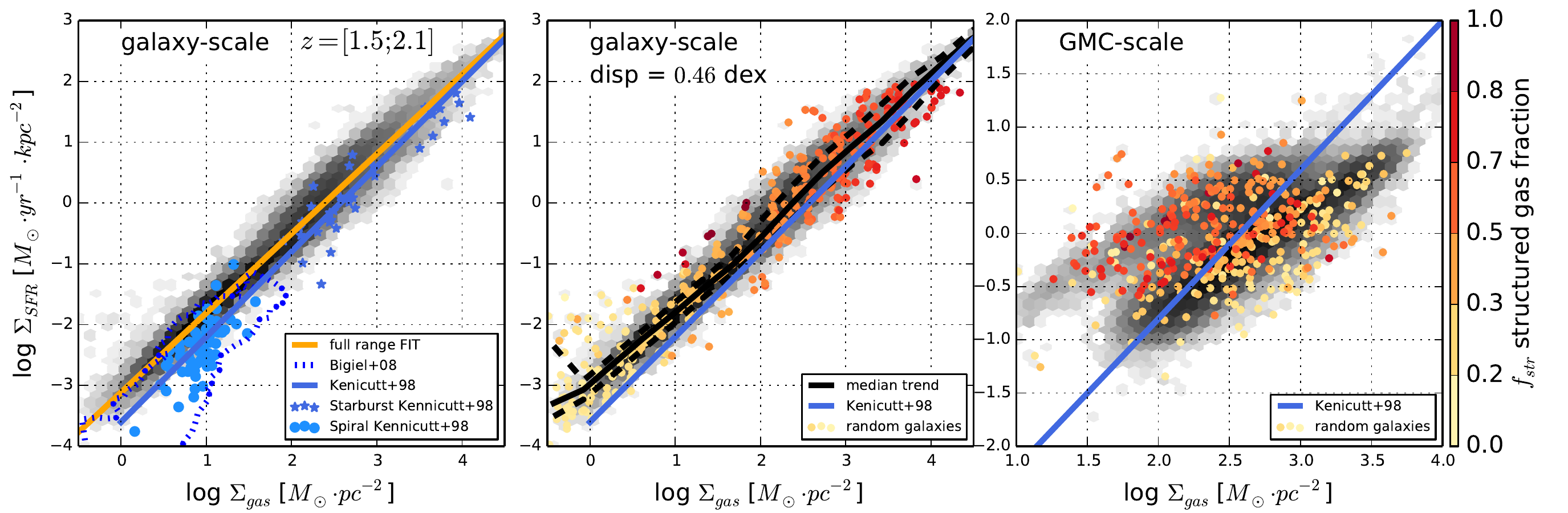}
  \caption{Schmidt-Kennicutt relations estimated from our model galaxies at both the galaxy-scale and GMC-scale. In the three panels the solid blue line indicates the \cite{Kennicutt_1998} relation. The grey shaded area is the distribution of a full sample of star-forming galaxies over the redshift range, $z$=1.5--2.1. Each coloured point indicates the position of an individual star-forming galaxy, uniformly selected within the galaxy-scale $\Sigma_{gas}-\Sigma_{SFR}$ plan. The colour bar indicates the fragmented gas fraction in each representative galaxy. \textit{Left panel:} The solid orange line is the best fit computed for our entire sample of star-forming galaxies within this redshift range. The dashed contour indicates measurements of nearby-galaxies \citep{Bigiel_2008}. Blue stars and points are from the starburst and normal disk galaxy samples from \cite{Kennicutt_1998}, respectively. \textit{Central panel:} The solid black line is the median trend of our star-forming galaxy sample. Dashed black lines indicate the 15\% and 85\% percentiles ranges of our model galaxies. We estimate an average scatter in the distribution of 0.46\,dex. \textit{Right panel:} The distribution of galaxies in the $\Sigma_{gas}-\Sigma_{SFR}$ estimated over the GMC-scale of star formation (see text for details).  We show this estimate for the same galaxy sample used in the galaxy-scale Schmidt-Kennicutt relation in the two leftmost plots in this figure.}
  \label{fig:SK_laws}
  \end{center}
\end{figure*}

\subsection{The fragmented gas reservoir}
\label{sec:fragmented_gas_reservoir}

The clumpy gas phase has low filling factor (typically less than $\approx 10$\%) with densities $n_\mathrm{H} = 1-10^3$~cm$^{-3}$ and temperatures 10$^3$-10~K. It is traced by atomic and molecular gas lines such as CO, [C{\sc i}], [C{\sc ii}] \citep[e.g.,][]{Aravena_2016, Bothwell_2016, Decarli_2016, Popping_2017, Bothwell_2017}. We assume that the fragmented gas is contained within spherical and bound structures with initial radii $r=h_d/2$, and must contain at least a mass $M_{BE}^{1/h}$ to be unstable. At each time step, the maximum number of bound structures formed in the disk is $N_{GMC}\approx(M_{frag}+M_{sfg})/M_{1/h_d}^{BE}$.

The gas fragmentation process feeds the star-forming gas reservoir at the rate:
\begin{equation}
	\dot{M}_{sfg}=\frac{M_{frag}}{\mathcal{T}_{frag}} \ .
	\label{eq:str_rate}
\end{equation}

$\mathcal{T}_{frag}$ is the effective disk fragmentation timescale. This timescale is updated at each time-step and allows us to follow the full history of the gas cycle in relation with the current disk properties ($h_d$ and $\dot{M}_{frag}$). If $M$ is the total mass of fragmented gas after the last time-step $\Delta t$, and $\Delta M = \Delta t \dot{M}_{frag}$ is the gas mass incorporated in the fragmented gas reservoir, thus the disk fragmentation timescale is updated as:
\begin{equation}
\mathcal{T}_{frag}^n=\mathcal{T}_{frag}^{n-1}\left(1-\frac{\Delta M}{M}\right)+\frac{\Delta M}{M}T_{frag}[h_d,\dot{M}_{frag}]  \ .
\label{eq:t_str}
\end{equation}

This timescale is a function of $T_{frag}[h_d,\dot{M}_{frag}]$, which depends on values of $h_d$ and $\dot{M}_{frag}$, the disk scale-height and the growth rate of GMCs, respectively (see Sect.~\ref{sec:struturing_time scale}).

\subsection{The star-forming gas reservoir}
\label{sec:star-forming_gas_reservoir}

Progressively, the fragmented gas is converted into star-forming gas. The gas contained in the
star-forming reservoir is very dense and cold, typically traced by molecular lines as HCN \citep[e.g.,][]{Oteo_2017}. The star-forming gas reservoir is characterised by its very short timescale before it forms stars. We assume a constant star-formation timescale which is linearly dependent on the length scale over which star formation occurs, $t_{sf} = l_{\star}/\sigma_{\star} = 0.1 Myr$. The star-formation rate is then simply given by:
\begin{equation}
	SFR = \dot{M}_{\star} = \dfrac{M_{sfg}}{t_{sf}}.
	\label{eq:star-formation_rate}
\end{equation}

In our prescription, the rate of star formation is mainly limited by the rate at which gas becomes clumpy.

\subsection{Schmidt-Kennicutt laws}

The \GAS\ model follows the evolution of three interacting gas reservoirs. Star formation is assumed to occur in the reservoir containing the densest gas after a progressive structuring starting at disk scale-height in some GMCs. We can estimate the Schmidt-Kennicutt laws at two different scales within this context.
At the galaxy scale, the gas surface density $\Sigma_{gas}$ is computed assuming that half of the mass of the fragmented gas is enclosed in the half mass radius (1.68$r_{d}$) of the disk. The star-formation rate surface density $\Sigma_{SFR}$ is computed in the same way. Using these two definitions, our predictions of the galaxy-scale Schmidt-Kennicutt law is shown in Fig. \ref{fig:SK_laws}. 

The median trend of our star forming galaxy sample at $z = 2.1$ is in good agreement with the \cite{Kennicutt_1998} relation over four orders of magnitude (Fig.~\ref{fig:SK_laws}). Below log$_{10}\Sigma_{gas}=1.0$, our star-formation rates are slightly higher than those deduced from \cite{Kennicutt_1998}. We measure an average scatter of $\simeq$0.46\,dex. Our star-forming galaxy distribution is fully consistent with individual measurements galaxies measurements \citep{Kennicutt_1998,Bigiel_2008}. Using the whole star-forming sample of galaxies over the redshift range $z$=1.5--2.1, we estimate a power-law index, $N = 1.232 \pm 0.002$. This slope is slightly shallower than the standard \cite{Kennicutt_1998} slope.
In addition, we find a clear trend for an increasing gas surface density as the fragmented gas fraction increases (Fig.~\ref{fig:SK_laws}).

In our model, we assume that star-formation is initiated in GMCs by the progressive fragmentation of the gas. GMCs are scaled to the disk scale-height $h_d$. By assuming that the fragmented gas mass and the star-formation rate is homogeneously distributed in all GMCs formed into the disk, we define the gas surface density and the star-formation rate surface density as follows:
\begin{equation}
	\Sigma_{gas} = \dfrac{1}{N_{GMC}}\dfrac{M_{frag}+M_{sfg}}{\pi (h_d/2)^2}~~\text{and}~~\Sigma_{SFR} = \dfrac{1}{N_{GMC}}\dfrac{SFR}{\pi (h_d/2)^2}
\label{eq:gas_sfr_surf_densities}    
\end{equation}

The clear correlation we found at the galaxy scale disappears at the GMC-scale (Fig.~\ref{fig:SK_laws}). Compared to galaxy scale, the fragmented gas fraction shows the opposite trend: the higher the gas surface density in a GMC, the lower its fragmented gas fraction. This trend can be translated as follow: galaxies with a relatively higher (lower) fraction of fragmented gas host relatively more (fewer) GMCs. The mass of gas and star-formation is homogeneously distributed in all GMCs (Eqs. \ref{eq:gas_sfr_surf_densities}). The average gas surface density is therefore lower in galaxies with highly fragmented gas which also happen to host more GMCs than galaxies with relatively low levels of fragmentation.

%
%

\section{Model Feedback}
\label{sec:feedbacks}

Massive stars and active galactic nuclei (AGN) in galaxies inject significant amounts of energy into the ISM and the circumgalactic medium (CGM) of galaxies. One of main challenges in galaxy evolution models is to distribute this power into the various gas phases in these media. We now describe how we distribute the SN and AGN power within the galaxy and its surroundings in the \GAS\ model, and how this power is used to regulate star formation.

\subsection{Morphology and the efficiency of outflows}
\label{sec:disc_morphology}

Galaxies in the early universe frequently have a clumpy morphology that suggest there are interacting regions of dense gas and stars \cite[e.g.,][]{Elmegreen_2009a, Elmegreen_2009b}. The clumps in these distant galaxies are not typically observed over the same range of mass and size as star forming regions in local disk galaxies. This morphological evolution provides clues as to how star formation in distant galaxies may have proceeded and was regulated, but overall observations are consistent with mass and energy injected into the ISM playing an important role in regulating star formation in galaxies \citep[self-regulation; e.g.,][and references therein]{Lehnert_2015}.

To account for this morphological evolution, we assign to each galaxy a specific morphological type. Galaxies that formed recently are assumed to be clumpy. The morphological type can then change, from clumpy to a smooth(er) disk, only during a merger. If two clumpy galaxies merge, the remnant galaxy can be converted to a smooth disk with a probability of 25\%. If two different morphologies merge, the remnant galaxy is assigned the morphology of the most massive progenitor (clumpy or smooth). Following these rules, clumpy galaxies become progressively regular, smoother disk galaxies. Furthermore, galaxies with these two different morphologies also behave differently as they evolve. For low mass, clumpy galaxies, we assume that gas is more compact and therefore that ejection of gas through outflows is more efficient. The terminal velocity of a wind is proportional to the square root of the energy injection rate divided by the total mass flow rate of the wind $\dot{M}_{wind}$. Thus, we can translate this efficiency in terms of the average terminal wind velocity. For clumpy galaxies, we assume the terminal velocity of the wind $V_{w} = 100$ \kms, and $V_{w} = 200$ \kms\ for disk galaxies. The mass loading factor $\dot{M}_{wind}$/SFR is higher in clumpy galaxies than in rotating smooth disks. At $z = 2.1$, distributions of the mass loading factor are characterised by the probabilities that they lie above the 25$^{th}$, 50$^{th}$ and 75$^{th}$ percentiles for all star-forming galaxies, which values are $\dot{M}_{wind}$/SFR= 6.5, 11.5, and 26.4, respectively.

\subsection{Supernovae feedback}
\label{sec:SN_feedback}

In \cite{Cousin_2016} the instantaneous SN event rate ($\eta_{SN}$ Gyr$^{-1}$) is proportional to the star-formation rate, which is a function of time (related to the star-formation history; see Sects.~3.5.2). Based on the rate of SNe, the instantaneous power generated by SN is simply given by $Q_{sn} = \eta_{sn}E_{sn}$. We assume the standard value $E_{sn} = 10^{44}$ erg s$^{-1}$ \citep[e.g.,][]{Aguirre_2001}. In \GAS, outside of the morphological division in the efficiency of outflows, we use the same prescriptions as in \cite{Cousin_2016}. Please see that paper for details.

\subsection{Active Nuclei feedback}
\label{sec:AGN_feedback}

In addition to the energy input from SNe, the different gas phases are also impacted by the energy produced by AGN. We assume that Super Massive Black Holes (SMBHs) are created during a major merger if the remnant galaxy have a bulge of at least $10^6\Msun$. The seed of the SMBH is given an initial mass, $M_{\bullet}^{init}=300\Msun$. We associate a gas torus to each SMBH formed. The torus is the gas reservoir that feeds the growth and energy output of the SMBH. This torus is fed by diffuse gas during merger events using the following prescription:
\begin{equation}
\Delta M_{torus} = 0.1\mu_g\mu_m M_{diff}(r<3r_{torus})
\label{eq:torus_mass_transfer}
\end{equation} 
where:
\begin{itemize}
	\item{$\mu_g$ is the mass fraction of the gas that is diffuse in the remnant galaxy disk;}
	\item{$\mu_m=\frac{\text{MIN}(M_1,M_2)}{\text{MAX}(M_1,M_2)}$ is the merger mass ratio and $M_i$ is the total mass (dark-matter + galaxy) enclosed in the half mass radius of the halo;}
	\item{$M_{diff}(r<3r_{torus})$ is the mass of diffuse gas enclosed in $r<3r_{torus}$. The torus radius, $r_{torus} = 10pc$.}
\end{itemize}

From Eq.~\ref{eq:torus_mass_transfer} it is clear that major mergers with gas-rich progenitors will accrete the largest amount of gas onto the torus of the remnant, while minor mergers of gas-poor progenitors will lead to very little accretion.
The infall rate is chosen to be the maximum value of the Bondi infall rate \citep{Bondi_1952} and the free-fall rate given by:
\begin{equation}
\dot{M}_{infall} = \text{MAX}\left(\frac{3\pi G \mu k_B T_{torus}}{4\Lambda(T_{torus},Z_{torus})}M_{\bullet};~\frac{M_{torus}}{2t_{\bullet}}\right),
\end{equation}
where the gas temperature of the torus is fixed to $T_{torus}=10^{6.5}$~K. $Z_{torus}$ is the gas-phase metallicity of the torus. The metallicity and temperature determine the cooling efficiency $\Lambda(T,Z)$ (see Sect.~\ref{sec:hot_mode}). $M_{\bullet}$ and $M_{torus}$ are the SMBH mass and the torus mass respectively. $t_{\bullet}$ is the orbital time of gas at the radius of the torus $r_{torus}$.

\begin{figure}[t!]
\begin{center}
\includegraphics[width=0.8\linewidth]{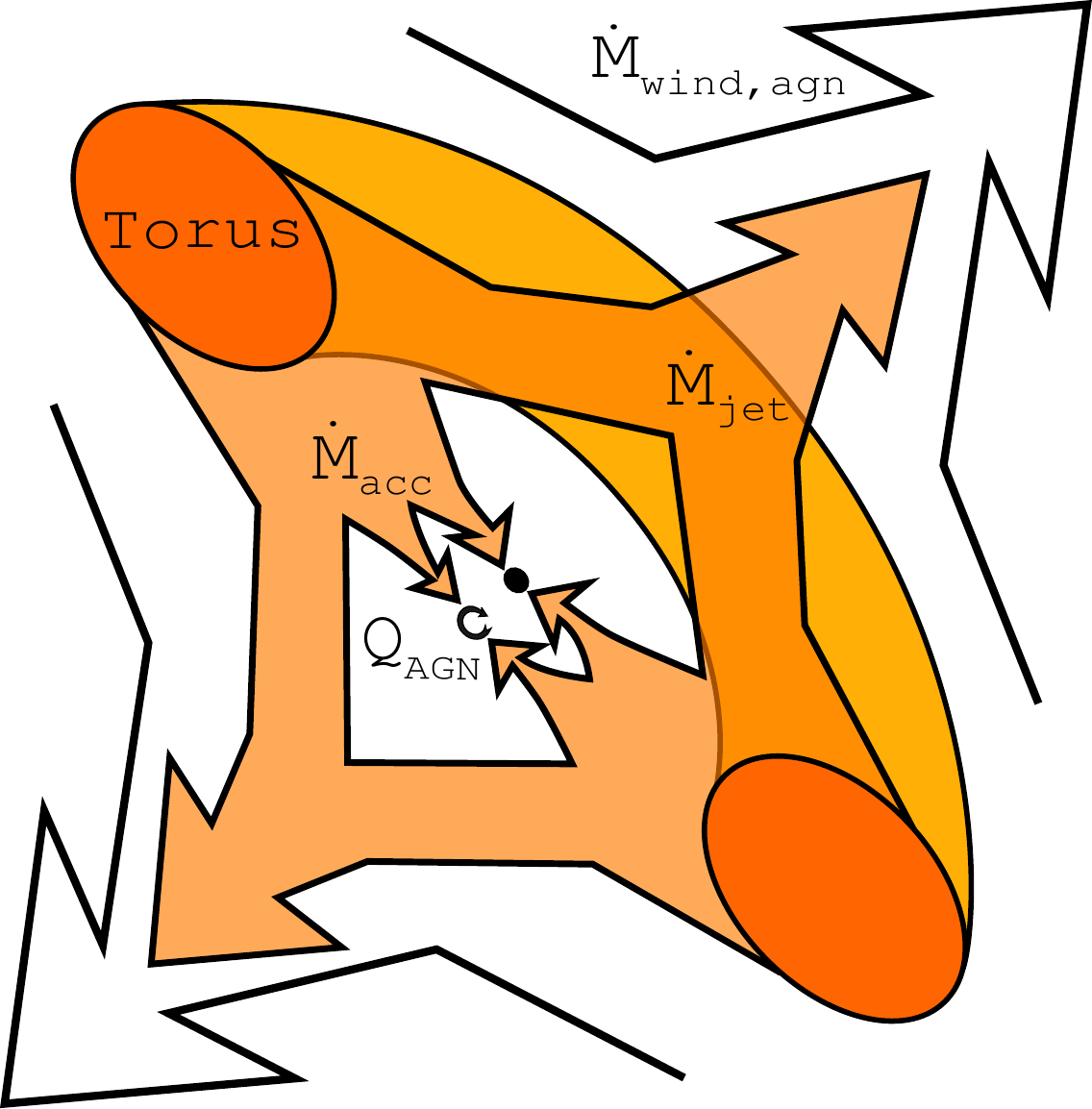}
\caption{\ftns{Illustration showing the distribution of accretion and energy output of a model AGN. Gas infalling from the torus to the AGN is divided between accretion onto the SMBH $\dot{M}_{acc}$, and ejection $\dot{M}_{jet}$. We assume the diffuse gas in the disk can be driven outwards by the AGN to give the overall ejection rate of gas $\dot{M}_{wind,agn}$. The accretion onto the SMBH $\dot{M}_{acc}$ is itself divided in two parts: a fraction $(1-f_{ME})\dot{M}_{acc}$ of the infalling gas goes to increasing the mass of the SMBH, while the remaining gas $Q_{agn} = f_{ME}\dot{M}_{acc}$ is converted into energy (see Sect.~\ref{sec:AGN_feedback} for details). The radiative and mechanical energy of the AGN is then distributed in the various ways as illustrated in Fig.~\ref{fig:feedback_energy_distribution}.}}
\label{fig:AGN_cartoon}
\end{center}
\end{figure}

Our prescription for the infall of gas on to the SMBH follows closely that presented in \cite{Ostriker_2010}. The total mass infall flux into the region of influence of the SMBH is the sum of the mass flux that is accreted onto the SMBH $\dot{M}_{acc}$, and the fraction that is driven out of the region of influence of the SMBH $\dot{M}_{ej}$. This yields the total infall rate $\dot{M}_{infall} = \dot{M}_{ej} + \dot{M}_{acc}$. In our model, we assume that the relative fraction of ejected to accreted mass flow rates is $\eta_{\bullet}=\frac{\dot{M}_{ej}}{\dot{M}_{acc}}=1.0$. This division results in equal shares of the infalling gas to be: \textit{(i)} accreted onto the SMBH, driving the increase the SMBH mass; and \textit{(ii)} generates power by converting a fraction of the accreted mass $f_{ME} = 0.1$ (Fig. \ref{fig:AGN_cartoon}). The power produced by the AGN is then $Q_{AGN} = f_{ME}\dot{M}_{acc}c^2$.

\subsection{Distribution of feedback power}

\begin{figure}[t!]
\begin{center}
\includegraphics[width=\linewidth]{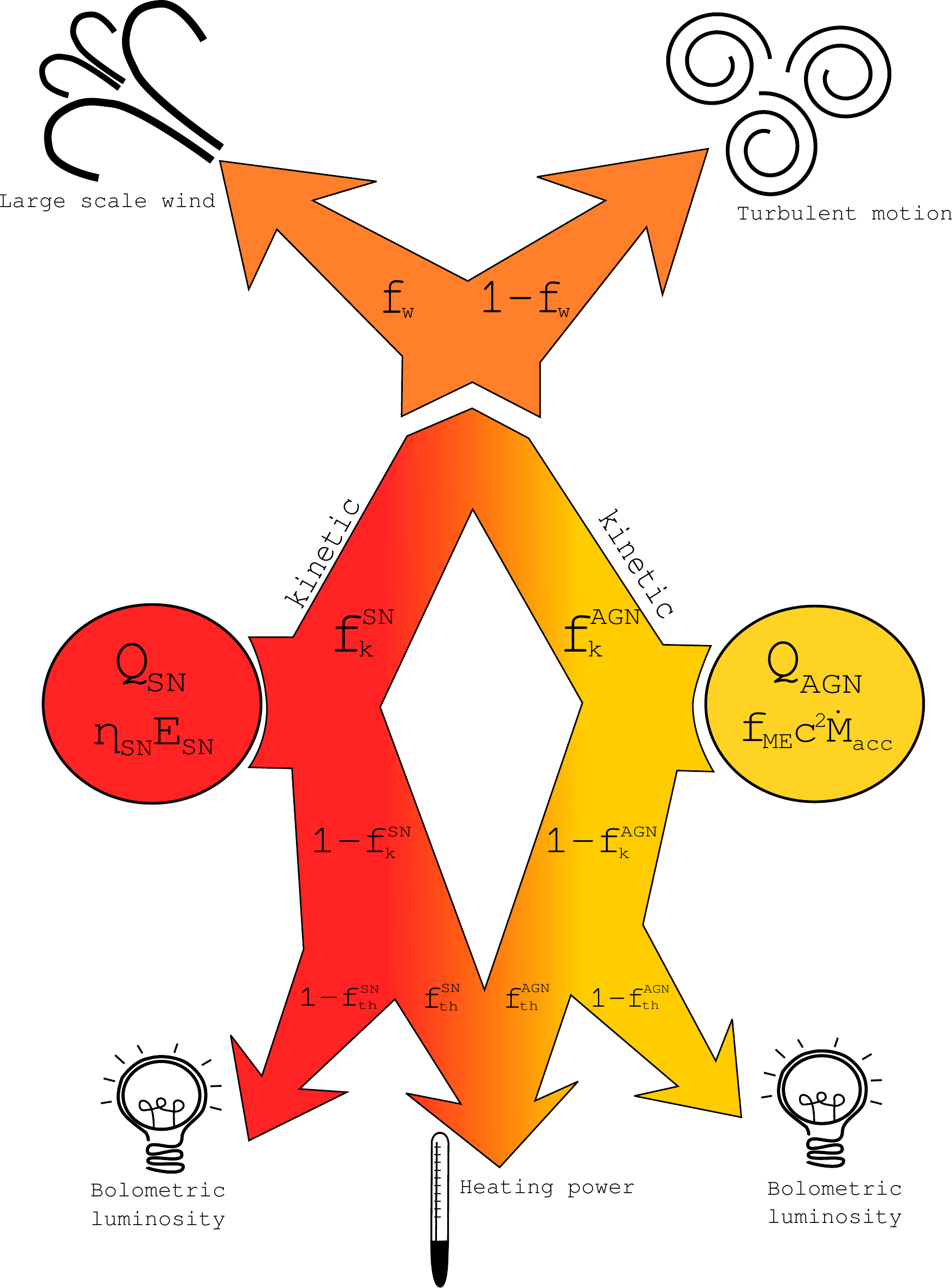}
\caption{\ftns{Illustration of how the energy output from AGN and SN are distributed. SN and AGN power $Q_{SN}$ and $Q_{AGN}$ are shown in red and yellow, respectively. Specifically, the power is distributed through four different channels: \textit{(1)} as a large scale wind; \textit{(2)} in generating turbulence; \textit{(3)} in heating the gas; and \textit{(4)} output bolometric luminosity.}}
 \label{fig:feedback_energy_distribution}
  \end{center}
\end{figure}

We use the power output from AGN and SN to regulate the gas cycle in galaxies in three different ways in our model. The two different sources of power --- the AGN power $Q_{AGN}$ and the mechanical energy of SN $Q_{SN}$ --- are each divided in two parts: kinetic power fraction $f_{k,AGN}$ and $f_{k,SN}$, and the bolometric power (Fig.~\ref{fig:feedback_energy_distribution}). A fraction of the kinetic power $f_w$ is used to drive a large scale wind (Sect.~\ref{sec:Large_scale_ejecta}). The residual fraction $1-f_w$ is used to disrupt the fragmented gas of the disk (Sect. \ref{sec:Disruption_rate}) and power the turbulence of the diffuse gas. A fraction of both the AGN and starburst bolometric luminosity, $f_{th,AGN}$ and $f_{th,SN}$ respectively, are used to heat the ejected gas (Sect.~\ref{sec:thermal_luminosity_power}).

\subsubsection{Large scale ejecta}
\label{sec:Large_scale_ejecta}

We derive the instantaneous ejection rate using the conservation of the kinetic energy released by SNs \citep[e.g.,][]{Dekel_1986, Kauffmann_1993, Efstathiou_2000}:
\begin{equation}
\dot{M}_{wind,SN} = 2f_wf_{k,SN}f_{esc}\dfrac{Q_{SN}}{v_w^2}
\label{SN_ejecta_rate_EQ}
\end{equation}
where $v_w$ is the average velocity of large scale wind (Sect.~\ref{sec:disc_morphology}). $f_{esc}$ is the fraction of mass that escapes the disk. 

The ejection rate due to the energy output of the AGN is determined by the fraction of the mass infalling towards the SMBH driven out ($\dot{M}_{ej}$; Sect.~\ref{sec:AGN_feedback}). We derive the velocity of the jet using conservation of energy:
\begin{equation}
v_{jet}^2 = \dfrac{2f_wf_{k,AGN}Q_{AGN}}{\dot{M}_{ej}}=\dfrac{2f_wf_{k,AGN}f_{ME}}{\eta_{\bullet}}c^2
\label{eq:Vjet}
\end{equation}

The outflowing jet is coupled to the diffuse gas in the disk which reduces its velocity. To obtain the coupling efficiency between the jet and the diffuse gas, we assume that jet velocity is equal to the escape velocity of the galaxy $V_{esc}$. Thus, the kinetic energy of the jet is simply used to drive the diffuse gas out of the galaxy. This leads to an mass outflow rate given by:
\begin{equation}
\dot{M}_{wind,AGN} = \dot{M}_{ej}\text{MAX}\left[1.0,\left(\dfrac{v_{jet}}{V_{esc}}\right)^2\right]
\label{eq:AGN_ejecta_rate}
\end{equation}

The total instantaneous ejection rate of the combined action of SN and AGN is $\dot{M}_{wind} = \dot{M}_{wind,SN} + \dot{M}_{wind,AGN}$. Each gas reservoir, $M_{diff}$, $M_{frag}$ and $M_{sf}$, contributes to the instantaneous ejection rate in proportion of its mass fraction (Fig.~\ref{fig:disc_structure}). 

\subsubsection{Overestimating the gas escape fraction: Re-accretion timescale}
\label{sec:2-orbits_re-accretion}

When the gas is ejected from the disc, a fraction of the gas will remain in the hot circumgalactic medium. The remaining fraction is ejected from the dark matter potential. As in \citet[][see their Sect. 4.3 and Fig. 6]{Cousin_2015b} to compute the escape fraction of the gas, we adopt a ``ballistic'' approach based on the comparison of the dark-matter escape fraction to a shifted Maxwell-Boltzmann velocity distributions. However, as shown by, e.g., \cite{Oppenheimer_2008}, the ejected gas is not only affected by gravitational forces ($V_{esc}$) but also by ram pressure of gas in the CGM. Due to this affect, our ballistic approach overestimates the escape fraction. To correct this bias, we store all the hot gas ejected from a halo in a extended circumgalactic reservoir $M_{ex-cir}$. The mass stored in this reservoir is then progressively re-accreted and added to the hot gas trapped into the dark-matter potential well. 

The gas that is re-accreted out of this reservoir has been implemented in various semi-analytical models \citep[e.g.,][]{DeLucia_2004b, Somerville_2008, Guo_2011, Henriques_2013} and potentially plays a major role in contributing to the overall gas supply of galaxies. The formulation in the various semi-analytical models can vary. For example, \cite{Guo_2011} adopt a prescription depending on both of the dark-matter halo mass and redshift. In \citep{Henriques_2013} the re-accretion of gas is inversely proportional to the dark-matter virial mass without any dependence on redshift. In \GAS, we adopt a prescription similar to \cite{DeLucia_2004b} and \cite{Somerville_2008}. We assume that the gas is re-accreted on a timescale which is twice the halo crossing time $t_{reacc} = 2r_{vir}/v_{vir}$.

\subsubsection{Disruption rate}
\label{sec:Disruption_rate}

The power generated by SNs that does not contribute to driving a wind is assumed to be injected directly into the ISM. This remaining power is distributed between the different ISM gas phases in proportion to their mass. Massive stars are assumed to remain mostly embedded in their dense birth clouds during their short lifetimes \citep[$3\times 10^6 yr$;][]{Cousin_2016}. The fraction of the SN energy which is injected into the fragmented and star-forming gas $f_{frag}$ disrupts this phase, feeding the diffuse gas reservoir (Eq.~\ref{eq:fragmented_gas_fraction} and Fig.~\ref{fig:disc_structure}). The disruption rate is defined as:
\begin{equation}
	\dot{M}_{disrupt,SN} = 2f_{frag}f_{k}^{SN}\underbrace{\left[(1-f_w)+f_w(1-f_{esc})\right]}_{1-f_wf_{esc}}\dfrac{Q_{SN}}{\sigma_v^2}
	\label{eq:SN_disruption_rate}
\end{equation}
where $\sigma_v$ is the average velocity dispersion of the diffuse gas (Sect.~\ref{sec:average_velocity_dispersion}). We assume that the velocity dispersion of the disrupted gas is equal to that of the diffuse gas. Assuming this implies that if the diffuse gas is highly turbulent (high $\sigma_v$) then it takes more energy to disrupt the fragmented gas. The disruption rate depends on two terms specifically related to the gas phase. The first term, $1-f_w$, corresponds to the minimum fraction of SN power which disrupts the fragmented gas. The second term, $f_w(1-f_{esc})$, corresponds to the fraction of power that remains in the gas because of the limit imposed by the galaxy escape fraction. The fraction of power which does not contribute to the wind is therefore re-injected to disrupt the fragmented gas. This fraction increases with the galaxy mass. The fraction of the SNs power injected into the diffuse gas phase maintains or increases the level of turbulence of the diffuse gas (i.e. $\sigma_v$). During a time step $\Delta t$, the possible increase of the turbulent energy of the diffuse gas is given by $\Delta E_{\sigma}^{SN} = (1-f_{frag})f_{k}^{SN}(1-f_wf_{esc})Q_{SN}\Delta t$ (see Sect.~\ref{sec:average_velocity_dispersion} for all other contributions).

Simultaneously, we also include the energy output from AGN in disrupting the gas. The contribution from any AGN to the disruption rate is given by:
 \begin{equation}
	\dot{M}_{disrupt,AGN} = 2f_{frag}f_{k,AGN}(1-f_w)\dfrac{Q_{AGN}}{\sigma_v^2}
	\label{eq:AGN_disruption_rate}
\end{equation}

As for SNs, the residual power $1-f_{frag}$ is injected into the diffuse gas as turbulent energy: $\Delta E_{\sigma}^{AGN} = (1-f_{frag})f_{k,AGN}(1-f_w)Q_{AGN}\Delta t$. The total mass disruption rate is the sum of the SN and AGN contributions, i.e. $\dot{M}_{disrupt} = \dot{M}_{disrupt,SN} + \dot{M}_{disrupt,AGN}$.

\subsubsection{Radiative heating and bolometric luminosity}
\label{sec:thermal_luminosity_power}

In the previous two sections we discussed our prescriptions for the kinetic power of SN and AGN. The non-kinetic fraction of the total power is also divided in two different parts. A fraction $f_{th,SN}$ ($f_{th,AGN}$) of the non-kinetic SN (AGN) power is used to heat the ejected gas. These fractions are adjusted so that the average temperature of the ejected gas is between $10^{6}$ and $10^{7}K$. The residual power $(1-f_{k,SN})(1-f_{th,SN})$ for SNs is then assumed to be emitted as bolometric luminosity in each galaxy (i.e. the part of the total radiative power that does not go directly into heating the gas).

\subsection{Average velocity dispersion and disk scale height}
\label{sec:velocity_disperesion_disc_scale_height}

\subsubsection{Average velocity dispersion}
\label{sec:average_velocity_dispersion}

The gas disruption rates, as given in Eqs. \ref{eq:SN_disruption_rate}, \ref{eq:AGN_disruption_rate}, depend on the average velocity dispersion of the diffuse gas, $\sigma_v$. We compute and continuously update the velocity dispersion by taking into account simultaneously the kinetic energy injected by gas accretion, SNs, and AGN.

Our prescription is based on the evolution of the ``turbulent'' kinetic energy budget $E_{\sigma_v}$: 
\begin{equation}
	2E_{\sigma_v}=M_{diff}\sigma_v^2 \ ,
	\label{eq:Kinetic_energy_budget}
\end{equation}
$\sigma_v$ being the 3D gas velocity dispersion. Between two time steps, we assume that:
\begin{itemize}
	\item{A fraction, $f_{disp}=1/2$, of the turbulent energy is dissipated per orbital time.}
	
	\item{The turbulent energy is increased by $2\Delta E_{\sigma_v}^{acc} = f_{incr}\Delta M v_{acc}^2$, corresponding to a fractional increase, $f_{incr}$=1/3, per dynamical time, of the kinetic rotational energy of the newly accreted diffuse gas, $\Delta M$. We assume that the freshly-accreted diffuse gas ($\Delta M$) forms a thin rotating disk \citep[e.g.,][]{Mo_1998}. This thin gas disk merges with the pre-existing disk hosting a mass, $M_{diff}$, of diffuse gas. $v_{acc}$ is the orbital velocity of the freshly-accreted diffuse gas computed at the half mass radius.}
	
    \item{The total budget of the turbulent energy is finally increased by the energy injected in the diffuse gas by SNs and AGN, $\Delta E_{\sigma_v}^{SN}$ and $\Delta E_{\sigma_v}^{AGN}$ (see Sect. \ref{sec:Disruption_rate}).}
    
\end{itemize}

Considering all of these energy terms, after a time step $\Delta t$, we have,

\begin{equation}
	E_{\sigma_v}^{n+1} = \left[1-f_{disp}\left(\frac{\Delta t}{t_{dyn}}\right)\right]E_{\sigma_v}^{n} + \Delta E_{\sigma_v}^{acc}\left(\frac{\Delta t}{t_{dyn}}\right) + \Delta E_{\sigma_v}^{SN} + \Delta E_{\sigma_v}^{AGN}
	\label{eq:Updated_kinetec_energy_budget}
\end{equation}

\begin{figure}[t!]
  \begin{center}
  \includegraphics[width=0.98\linewidth]{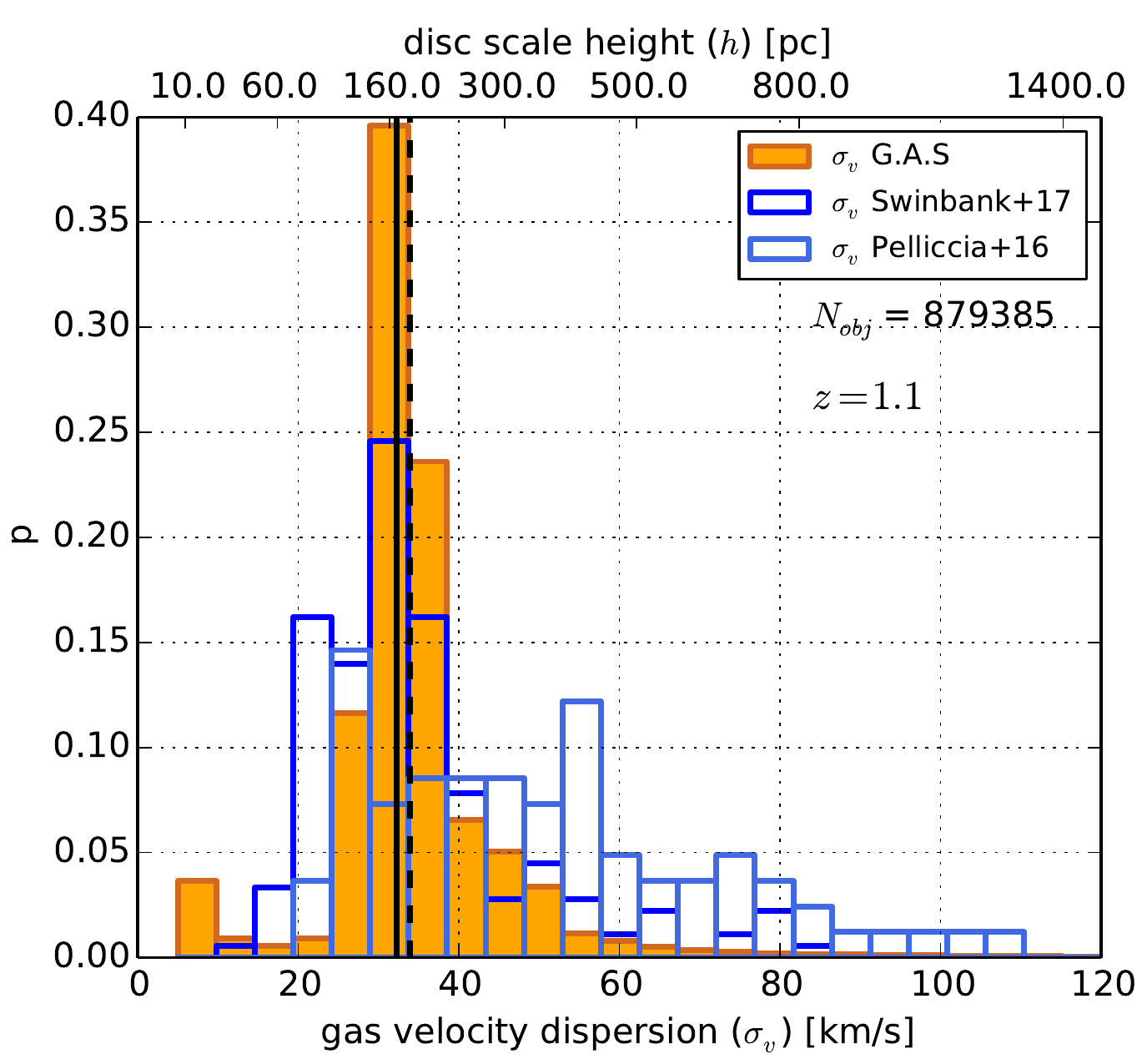}
  \caption{\ftns{Probability distribution of the velocity dispersion of the diffuse gas. This distribution is for model galaxies with $M_{\star}>10^{7}M_{\odot}$ and at $z=1.1$. We compare this distribution with observational measurements from \citet[][dark blue histrogram]{Swinbank_2017} and \citet[][light blue histrogram]{Pelliccia_2017}. The solid and dashed vertical lines represents the median and the mean value of the distribution of our modelled galaxies respectively.}}
  \label{fig:disc_scale_height_disc_velocity_dispersion}
  \end{center}
\end{figure}

Based on this updated total turbulent kinetic energy, we calculate the average velocity dispersion of the diffuse gas $\sigma_v$ (Eq. \ref{eq:Kinetic_energy_budget}). The two free parameters, $f_{incr}$ and $f_{disp}$, have been adjusted to reproduced observed values of velocity dispersion (see Fig.~\ref{fig:disc_scale_height_disc_velocity_dispersion}).

During a merger, the total turbulent kinetic energy of the two progenitors are added. We also add a fraction of the gravitational energy due to the interaction between the two galaxies,

\begin{equation}
	E_{int} = G\dfrac{M_1M_2}{1.68(r_d^1 + r_d^2)}
	\label{eq:Kinetic_energy_budget_merger}
\end{equation}
\noindent
where $M_i$ is the mass (baryon + dark-matter) included in the galaxy half mass radius, and $r_d^i$ is the disk exponential radius \citep{Hatton_2003}.

We compare our predictions of the gas velocity dispersion with recent observational measurements of the velocity dispersion of the warm ionised media from \cite{Swinbank_2017} and \cite{Pelliccia_2017}. Objects targeted by these two programs are distributed in redshift between $z \simeq 0.3$ and $z \simeq 1.7$. Our predictions are in good agreement with these observational results but our results to do reproduce some of the most extreme dispersions observed (both small and large values; Fig. \ref{fig:disc_scale_height_disc_velocity_dispersion}). At $z = 1.1$, the median value of the velocity distribution of the diffuse gas is close to 35 km s$^{-1}$.  

\subsubsection{Disk scale height}
\label{hd_appendix}
The disk fragmenting timescale (Eq.~\ref{eq:t_str}) depends on the average disk scale height. We use the disk scale height to define the initial energy injection scale of the inertial turbulent cascade. Our gas fragmenting scenario is mainly based on scaling relations and the disk scale height is crucial in calculating this parameters of the turbulent cascade. The disk scale height $h_d$ is deduced from $\sigma_v = \sigma(h_d^{-1})$.  
At $z = 1.1$, the median value of the velocity dispersion leads to a median value of the disk scale height is $\approx$160 pc (Fig. \ref{fig:disc_scale_height_disc_velocity_dispersion}).

%
%

\section{Thermal instabilities}
\label{sec:thermal_instabilities}

For galaxies with large masses, previous semi-analytical models invoked powerful AGN feedback to greatly reduce or even completely quench accretion of cooling gas in their halos \citep[e.g.,][]{Cattaneo_2006, Somerville_2008, Guo_2011, Benson_2012, Henriques_2013}. Such prescriptions assume a strong coupling between AGN power, the ISM, and the hot circum-galactic gas, which results in powerful outflows expelling the gas and heating the halo gas. However, observations are not entirely clear on the precise impact of outflows on distant galaxies\citep[see, e.g.,][]{Mullaney_2015, Netzer_2014, Netzer_2016, Scholtz_2018, Falkendal_2018}. In \cite{Cousin_2015a, Cousin_2015b}, we also assume that AGN can have an impact on the hot halo phase but we have limited the effect to simply heating the halo gas. 

In the following we present a new prescription to efficiently reduce the gas accretion onto massive galaxies, $M_{\star} > 10^{11}\Msun$. This new mechanism is based on the growth of thermal instabilities in the hot halo phase surrounding the galaxy.

\subsection{Thermal instabilities: Description}

Our new mechanism assumes that the condensation of gas and therefore the gas accretion onto the galaxy is progressively and strongly limited by the growth of thermal instabilities in the hot gaseous circum-galactic medium. In the standard model of cooling presented previously and also used in e.g., \citet{Croton_2006, Baugh_2006, Somerville_2008}, there is a strong dichotomy between the gas stored within and beyond the cooling radius, $r_{cool}$. At radii greater than the cooling radius, the gas is assumed to remain hot, while below $r_{cool}$, the gas is assumed to be warm, $10^4$~K, and dense enough to condense and feed the galaxy. Of course, the reality is more complex, and obviously hot and warm gas co-exist around the cooling radius. In this transition region, the gas can be thermally unstable. These thermal instabilities can generate warm clouds that are orbiting in the hotter gas. \cite{Cornuault_2018}, in a phenomenological model of accreting gas, find that the cloud-cloud velocity dispersion can be comparable to the characteristic virial velocity of the dark-matter halo in which they formed. The mixing and dynamics of warm clouds into the hot atmosphere can moderate the effective accretion rate of gas onto a galaxy.

\subsubsection{Thermal instability clock}
\label{sec:TI_clock}

As shown by \cite{Sharma_2012}, the timescale over which gas becomes thermally unstable is related to the cooling time. Following their approach, and similar to gas cooling generally, we define a thermal instability timescale, $\mathcal{T}_{TI}$. This timescales evolves concomitantly with the cooling timescale. After each time step, $\Delta t$, the thermal instability clock is updated following a mass-weighted prescription, similar to the one applied to the cooling clock, namely,

\begin{equation}
   \mathcal{T}_{TI}^n = \underbrace{\left(\mathcal{T}_{TI}^{n-1} + \Delta t_{TI}\right)\left(1-\frac{\Delta M}{M}\right)}_{\text{hot gas in the halo}} + \underbrace{\frac{\Delta t_{TI}}{2}\frac{\Delta M}{M}}_{\text{freshly accreted hot gas}}
   \label{eq:TI_clock}
\end{equation}

\noindent
where $M$ is the total mass stored in the hot phase after the last time step, $\Delta t$. $\Delta M$ is the mass accreted by the hot gas phase and/or ejected from the galaxy. For each time step, $\Delta t$, the time increment $\Delta t_{TI}$ is calculated as,

\begin{equation}
	\Delta t_{TI} = 0.1F(\Lambda)\Delta t
\end{equation}

\noindent
where $F(\Lambda)$ is the thermal instability function \citep{Sharma_2012}, depending on the logarithmic derivative of the cooling efficiency function $\Lambda$,

\begin{equation}
   F(\Lambda) = 2-\dfrac{dln\Lambda}{dlnT}
   \label{eq:F_Lambda_TI}
\end{equation}

We assume that the hot gas becomes thermally unstable during all the previous time steps (Eq.~\ref{eq:TI_clock}). However, as for the effective cooling clock, we consider that thermal instabilities can evolve with the addition of freshly accreted gas only during half a time step\footnote{Following a scheme where gas is continuously added into the hot gas reservoir.}.
The thermal instability function, $F(\Lambda)$, determines if the hot gas is stable or unstable. Gas is assumed to be unstable if $F(\Lambda) \ge 0$ and stable if $F(\Lambda) < 0$. When hot gas is stable, the time increment $\Delta t_{TI}$ is set to 0. According to the pre-existing/incoming mass ratio, the thermal instability clock can increase or decrease during a given time step.

\begin{figure}[t!]
    \centering
        \includegraphics[width=\linewidth]{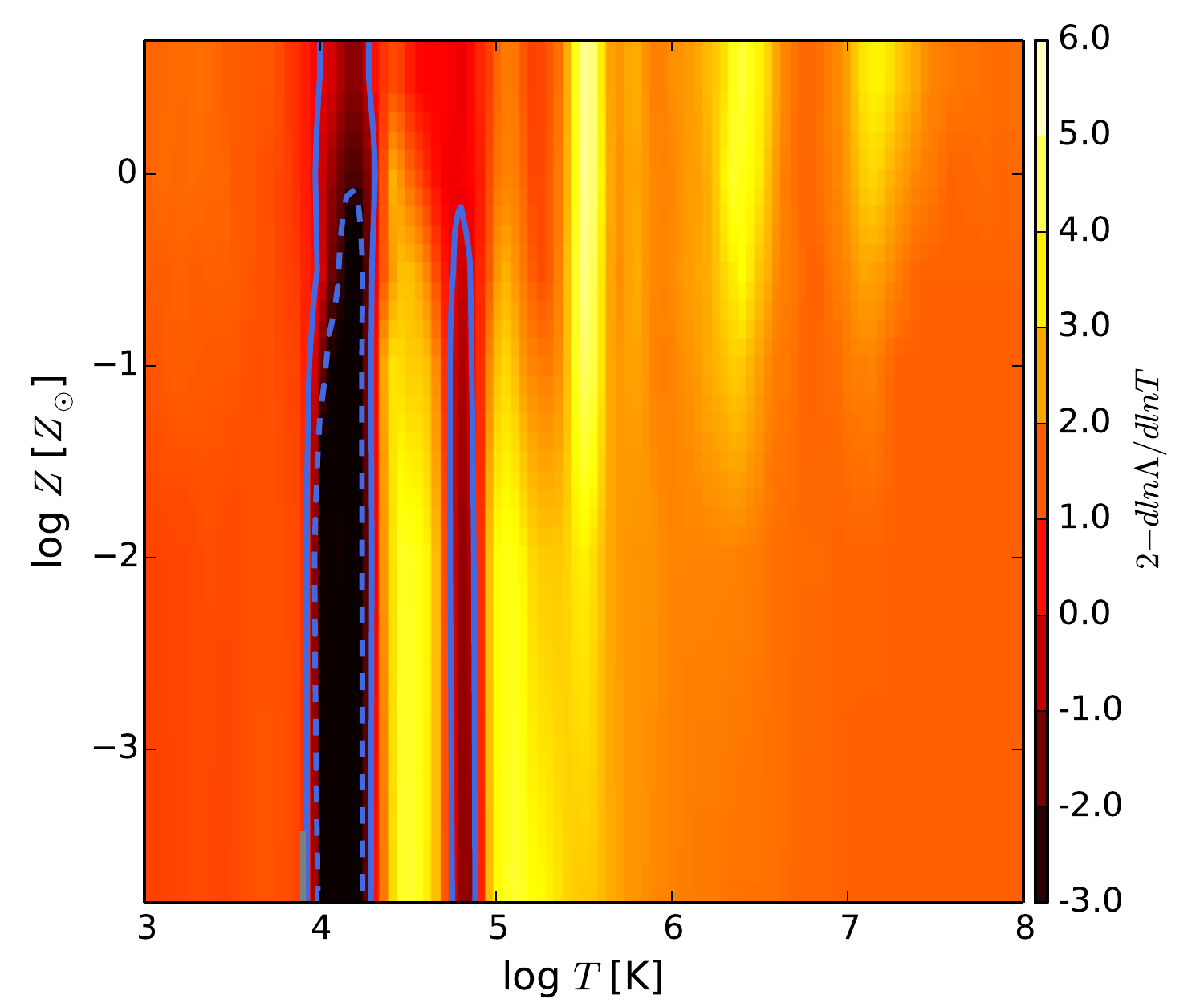}
    \caption{Thermal instability efficiency, $F(\Lambda)$, as a function of both gas temperature and gas metallicity. The amplitudes of thermal instabilities are computed as given in \cite{Sharma_2012} and the values are indicated in the color bar on the right side of the plot. When $F(\Lambda) < 0$, the gas is thermally unstable. The solid and the dashed blue contours indicate where $2-dln\Lambda/dlnT$=0 and $-$3.0, respectivelly.}
    \label{fig:cooling_TI_eff}
\end{figure}

As for the cooling efficiency, the thermal instability function has been interpolated and tabulated. Figure~\ref{fig:cooling_TI_eff} shows the thermal instability function dependence on gas temperature between $10^3$K and $10^8$K and gas metallicity between $10^{-4}Z_{\odot}$ and $2Z_{\odot}$. As for the effective cooling clock, the thermal instability clock of the hot phase is assumed to have the value that the most massive progenitor had prior to merging.
 
\subsubsection{The mixing zone}
\label{sec:Mixing_zone}

The thermal instability clock provides at any time the effective timescale of the growth of thermal instabilities. Starting at the cooling radius, we assume that thermal instabilities are propagating through the gas and ultimately reach the center of the halo. We define the instantaneous size of the mixing zone as,

\begin{equation}
   \Delta r_{TI}^n = \varepsilon_{TI}t_{TI}^n\sqrt{\dfrac{\gamma R T_{hot}}{\mu}} \ .
   \label{eq:TI_radius}
\end{equation}

\noindent
We assume an adiabatic index $\gamma = 1.4$ and a mean molecular mass $\mu = 0.62$. $R$ is the specific ideal gas constant and $T_{hot}$ is the average temperature of the hot gas. We assume that the propagation speed of these instabilities is close to the sound speed of the gas. Following this hypothesis, the value of the free parameter $\varepsilon_{TI}$ is set to 0.63. 

In the mixing zone, we assume that thermal instabilities lead to the formation of warm gas clouds orbiting around the galaxy. As already stated, the cloud-cloud velocity dispersion is assumed to be close to the virial velocity \citep{Cornuault_2018}. In addition, even if we do not take into account explicitly such a mechanism, the warm gas clouds will probably interact strongly with the large-scale wind escaping the galaxy. Through this interaction, we can reasonably assume that kinetic energy of the outflowing gas is injected into a dynamic and cloudy circum-galactic medium. Thus we assumed that the condensation of warm clouds is suppressed in the mixing zone. The condensation of clouds is only efficient in the inner region of the hot atmosphere and the effective cooling rate is then computed using Eq.~\ref{eq:cooling_rate}. For this effective cooling rate, the initial cooling radius calculated is substituted with the expression, $r_{TI} = r_{cool} - \Delta r_{TI}$.  

\subsection{Impact of thermal instabilities on the gas cooling}
\label{sec:Impact_of_thermal_instabilities}

To understand and quantify the progression of thermal instabilities in the hot gas phase, we define the thermal instability volume filling factor as,

\begin{equation}
	\phi_v = 1.0 - \left(\frac{r_{TI}}{r_{cool}}\right)^3 \ .
	\label{eq:VFF}
\end{equation}

\noindent
$\phi_v$ measures the volume fraction in which thermal instabilities are developed. $\phi_v = 0$ indicates that there is no impact on the gas cooling. As in the original model, all the gas enclosed in the cooling radius can condense and therefore feed the galaxy. $\phi_v=1$ indicates that all the region with $r<r_{cool}$ is thermally unstable. When $\phi_v=1$, since the clouds are orbiting at the virial velocity, is the point at which the cooling flow would otherwise occur, is effectively stopped. 

\begin{figure}[t!]
  \begin{center}
  \includegraphics[width=0.98\linewidth]{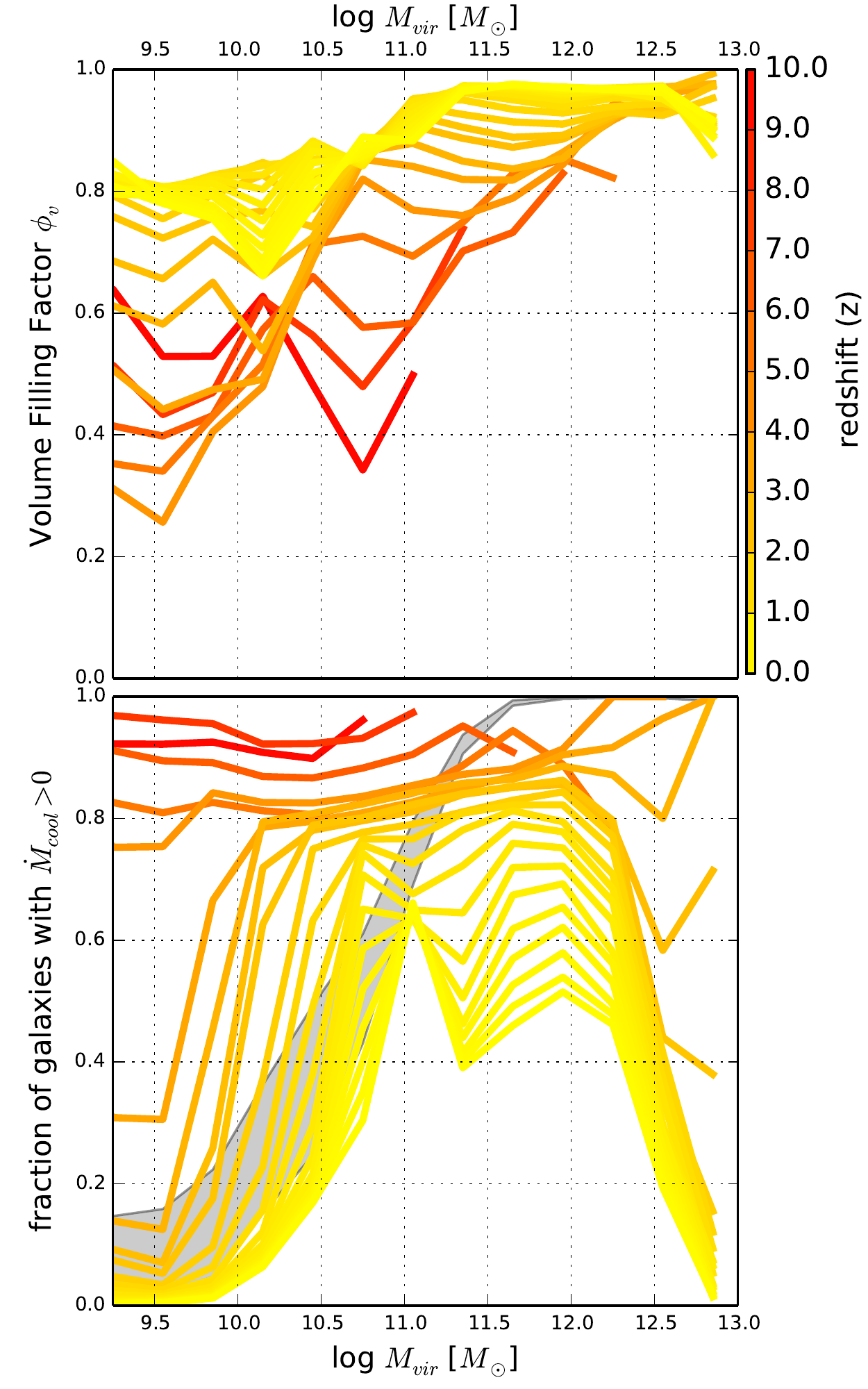}
  \caption{Upper panel: Mean value of the volume filling factor, $\phi$, of gas that is thermally unstable (Eq.~\ref{eq:VFF}). $\phi_v=1$ implies that all the region within $r = r_{cool}$ is thermally unstable in our model. Lower panel: The fraction of galaxies that are accreting radiatively cooling halo gas as a function of halo virial mass, M$_{vir}$. Different colours indicate this fraction as a function of redshift. The value of the redshift for each line is indicated in the colour bar on the right side of the upper panel. The gray area spans the range of the mean fraction of galaxies that are accreting hot gas for each halo mass summed over redshifts 0.1 to 10 (Eq.~\ref{eq:shock_heated_fraction}). For example, for a halo mass of 10$^{10.5}$ M$_{\sun}$, the fraction of galaxies that have hot gas in the halo which is radiatively cooling can range between 30\% to $\sim$55\% depending on redshift.}
  \label{fig:TI_VFF}
  \end{center}
\end{figure}

Fig. \ref{fig:TI_VFF} shows the mean trend of $\phi_v$ (Eq. \ref{eq:VFF}) as a function of both the dark-matter halo mass and redshift. Depending on the redshift, the first resolved halos,  M$_{vir}$>$10^{9}\Msun$, have $\phi_v$ distributed between 0.3 and 0.8. At $z \ge 9.0$, the average $\phi_v$ of the first halos formed is close to 0.6. This value progressively decreases with the redshift. At $z\simeq 4.0$, the average $\phi_v$ reaches a minimum around 0.3. Then the average $\phi_v$ of first halos detected increases as the redshift decreases. The maximum value, 0.8, is reached at low redshift. In this low mass regime, hot atmospheres are only formed via large-scale winds coming from the galaxy. The average $\phi_v$ is closely linked to galaxies driving winds. Indeed, as mentioned in Sect. \ref{sec:TI_clock}, the effective TI clock runs in proportion to the freshly added to pre-existing gas mass ratio in the hot halo phase (Eq. \ref{eq:TI_clock}). The higher the proportion of freshly added ejecta, the slower the TI clock advances and thus the slower $\phi_v$ increases. Between $z\simeq 9.0$ and $z= 4.0$, the intensity of ejecta affecting the first resolved halos, M$_{vir}\simeq 10^{9}\Msun$, increases continuously (related to both the gas accretion and the SFR). Then the intensity progressively decreases until the lowest redshifts. We caution that the number of halos in the low mass bins at very high redshift, $z \simeq 9$, is low and thus these bins are greatly affected by statistical noise. 

Over the redshift range, $z = 4.0$ to $z = 1.5$, the volume filling factor of the most massive halos reaches very high $\phi_v$, $\simeq 0.95$. At $z < 1.0$, the average volume filling factor of thermally unstable gas in massive halos starts to decrease. In these massive halos, some accretion still occurs. This accretion produces a new star formation and therefore some large scale ejection events. In these massive halos we therefore observe an increase of freshly acquired hot halo gas. This increase in freshly acquired hot halo gas results in a progressive decrease of the volume filling factor of thermally unstable gas.  

\begin{figure*}[t!]
  \begin{center}
  \includegraphics[width=\linewidth]{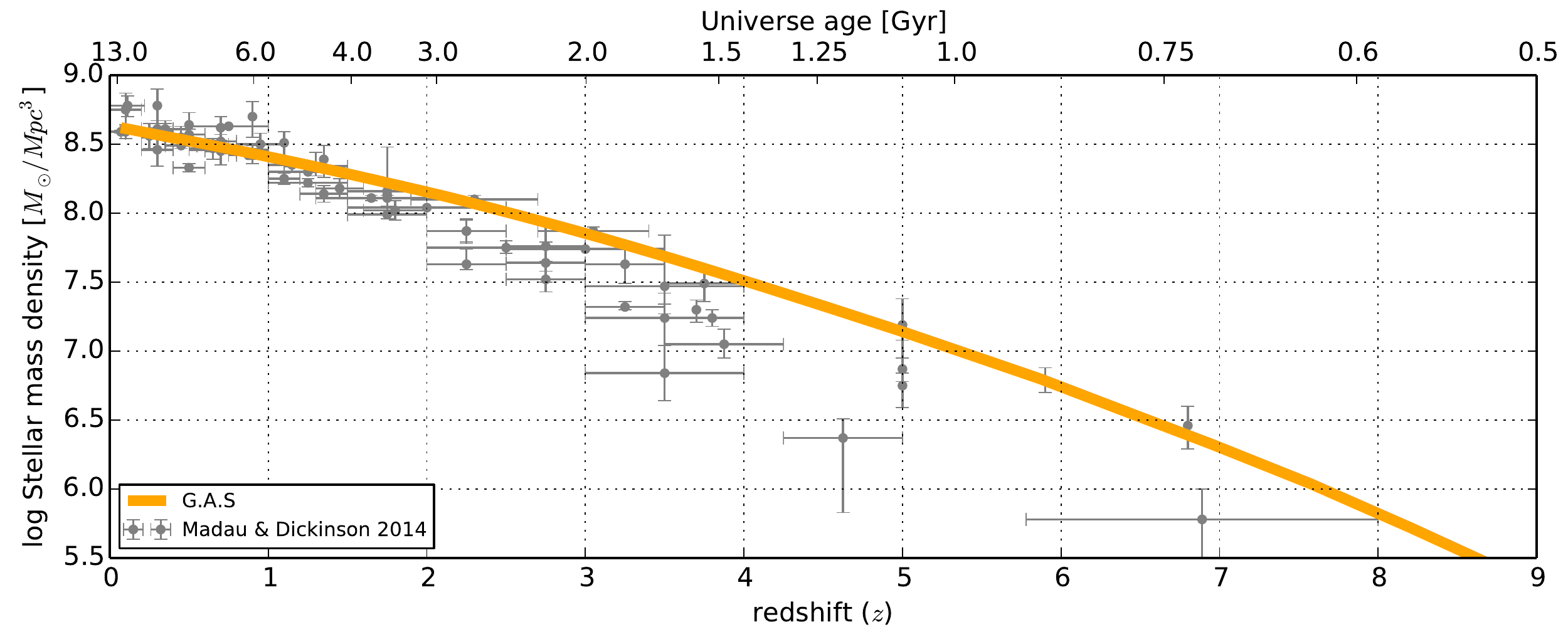}
  \caption{The redshift evolution of the co-moving stellar mass volume density. The solid orange line is the prediction of the \GAS\ model. We compare our predictions with the compilation of observational results \citep[see][and references therein]{Madau_Dickinson_2014}.}
  \label{fig:cosmic_stellar_mass_density}
  \end{center}
\end{figure*}

Fig. \ref{fig:TI_VFF} also shows the change the dark matter halo mass and the evolution with redshift of the fraction of galaxies in which condensation of the hot halo gas is occurring, i.e., $\dot{M}_{cool}>0.$ (Eq.~\ref{eq:cooling_rate}). At the halo detection mass threshold of our model, radiatively cooling gas is accreting over the whole mass range. The higher the redshift, the higher the fraction of galaxies that accrete radiatively cooled hot halo gas. For $z > 5.0$, the fraction remains high, $>0.8$. Then it strongly decreases and reaches values smaller than 0.1 at $z < 3.0$.
At low virial halo masses, the hot gas phase is only fed by large-scale ejecta from galactic outflows ($f_{hot} \le 0.18$). In low virial-mass halos, the mass of hot halo gas is small, $M_{hot}\le 10^{7}\Msun$. The cold mode is the dominant accretion mode at low masses. Even if at $z>5.0$ radiative cooling produces an effective accretion in a large fraction of the galaxies, the average radiative cooling rate is less than 1$\Msun/yr$ (Fig.~\ref{fig:acc_rate}). The fraction of dark matter halos hosting actively radiatively cooling hot gas increases (or remains constant) as a function of the dark-matter virial mass. The largest population of actively cooling hot halos is reached between $\simeq 10^{10.25}\Msun$ and $\simeq 10^{12}\Msun$. Between $z = 4.0$ and $z = 3.0$, in the most massive halos $>10^{12}\Msun$ the fraction reaches 1.0. However by $z < 3.0$ for the most massive halos, we find a strong decrease in the fraction of galaxies with actively cooling halos. This decrease is directly linked to the high $\phi_v$ reached in these halos and this decrease is responsible for the reduction in the effective cooling rate estimated for the most massive halos at $z<2$ (Fig. \ref{fig:acc_rate}).

%
%

\section{Results and analysis}
\label{results}

\subsection{The stellar mass assembly and co-moving density of galaxies}

\begin{figure*}[t!]
  \begin{center}
  \includegraphics[width=\linewidth]{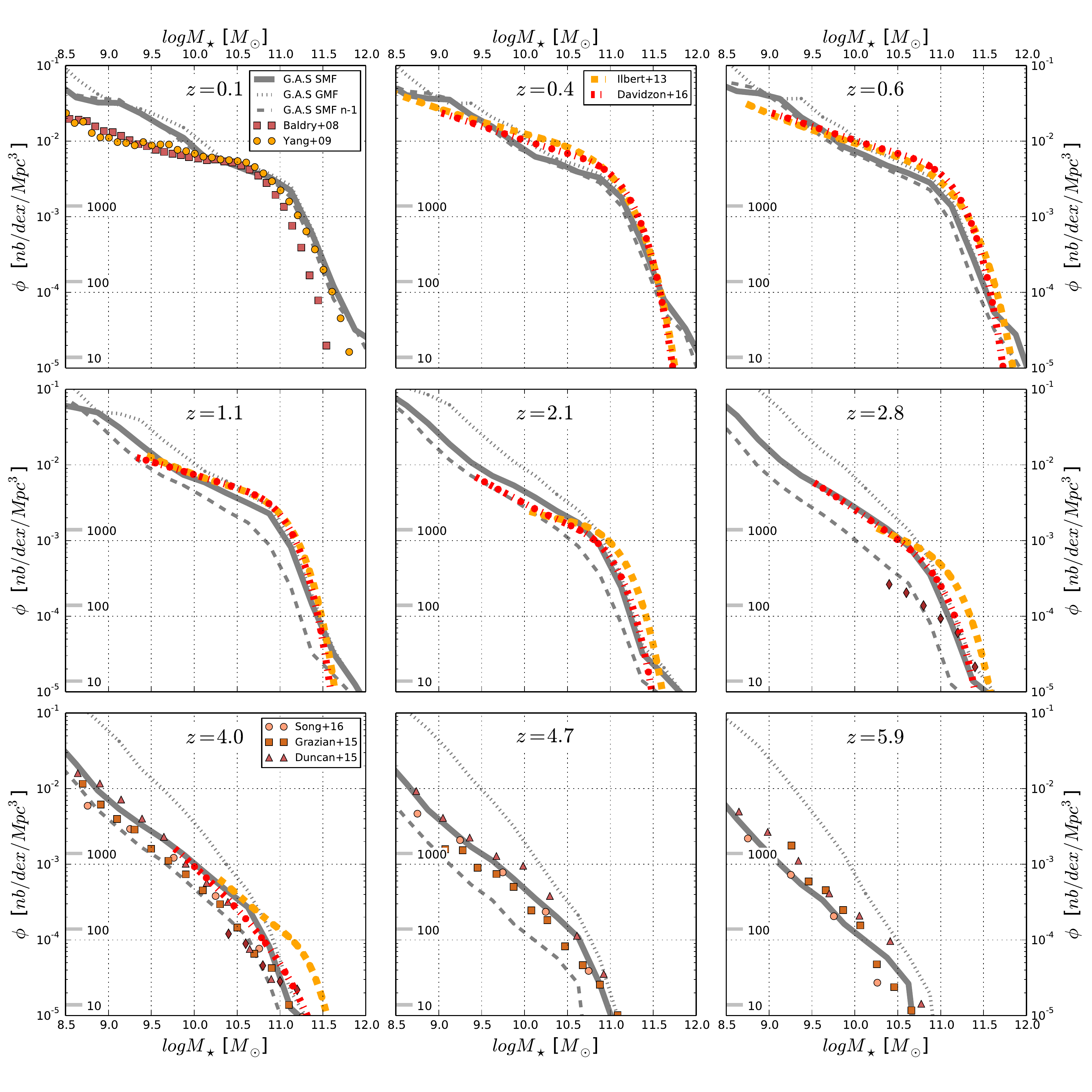}
  \caption{A comparison between the predictions of \GAS\ with observational estimates of the co-moving number density of galaxies for 9 redshift bins spanning the range, $z\simeq 0.1$ to $z\simeq 6$ (from top left to bottom right and as indicated at the top of each panel). We compare the predictions of the \GAS\ model (solid grey lines) with various studies: \cite[][red squares]{Baldry_2008}, \cite[][orange circles]{Yang_2009}, \cite[][brown diamonds]{Caputi_2011}, \cite[][chocolate squares]{Grazian_2015}, \cite[][indian-red triangles]{Duncan_2015}, \cite[][salmon circles]{Song_2016}, \cite[][yellow dashed-line]{Ilbert_2013}, and \cite[][red dot-dashed line]{Davidzon_2016}. For this comparison with observations, we indicate binned measurements using points and best fit stellar mass functions as dashed lines (Schechter and double-Schechter). All stellar mass functions are for a Chabrier IMF. The best fits of \cite{Ilbert_2013} and \cite{Davidzon_2016} are corrected for the Eddington bias which affects the high-mass galaxy bins in particular. The gray dotted-line indicates the total, gas plus stars, baryonic mass function predicted by \GAS. The gray dashed-line indicates the stellar mass function of the previous panel for the next highest redshift bin. The tick marks placed on the left side of each panel indicates the number of galaxies used in the stellar mass function predicted by the \GAS\ model. }
  \label{fig:stellar_mass_function}
  \end{center}
\end{figure*}

\begin{figure*}[t!]
  \begin{center}
  \includegraphics[width=\linewidth]{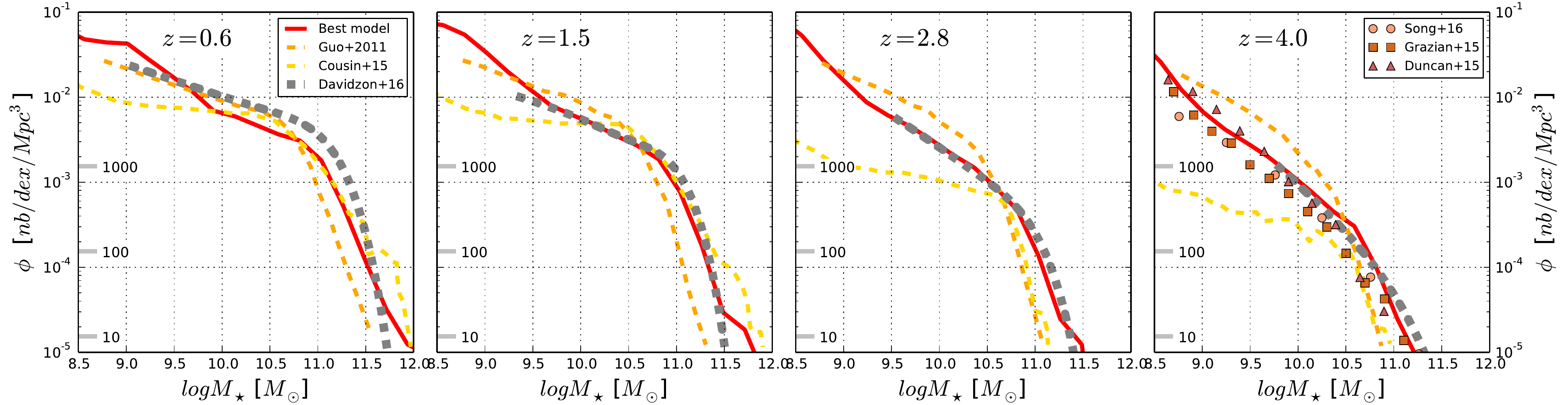}
  \caption{Comparison of the stellar mass functions predicted by our best fiducial model with those predicted by \cite{Guo_2011}. In each panel, each of which are for different redshifts as indicated at the top left in each panel, solid red, dashed orange and dashed gold lines indicate the predictions of \GAS, \cite{Guo_2011}, and \cite{Cousin_2015a}, respectively (as indicated in the legend of the left-most panel). We compare these models with observations \citep[dashed grey lines;][]{Davidzon_2016}. At $z=4.0$, we also plot additional observational measurements \citep[salmon circles, red upward pointing triangles, and orange squares;][respectively and as indicated in the legend of the right-most panel]{Song_2016,Duncan_2015,Grazian_2015}}
  \label{fig:stellar_mass_function_comparison_other_SAM}
  \end{center}
\end{figure*}

We compare the predictions of the \GAS\ model for the co-moving stellar mass density as a function of redshift with a compilation of observations (Fig.~\ref{fig:cosmic_stellar_mass_density}). Overall, the stellar mass growth of the Universe predicted in our model are in agreement with observations. Predictions are in reasonable agreement with estimates even at the highest redshifts, $z>3$, although generally consistent with the highest total co-moving stellar mass density estimates. 
Furthermore, this general agreement between \GAS\ and observations is also reflected in a comparison of the co-moving number density of galaxies binned by redshift (Fig.~\ref{fig:stellar_mass_function}). We compare our results with a variety of observed distributions covering the redshift range $z\simeq 0.1$ to $z \simeq 6$ \citep{Baldry_2008, Yang_2009, Caputi_2011, Ilbert_2013, Grazian_2015, Duncan_2015, Song_2016, Davidzon_2016}.
In addition, we compare our best \GAS\ predictions for the co-moving number density with those of \cite{Guo_2011} and also those obtained with the previous \textit{ad-hoc} regulation prescriptions of \cite{Cousin_2015a}.  These comparisons allow us to gauge the impact of including the star-formation and gas-dissipation prescriptions that are in \GAS\ with models that do not include such prescriptions (Fig.~\ref{fig:stellar_mass_function_comparison_other_SAM}).

Between $z\simeq1.0$ and $z\simeq6.0$, our model agrees very well with observational measurements of both the cosmic comoving mass densities and shape of the co-moving number density of galaxies for a wide range of redshifts (cf. Figs.~\ref{fig:cosmic_stellar_mass_density} and \ref{fig:stellar_mass_function}). At $z<1.0$, our model systematically under-predicts the comoving density of intermediate mass galaxies and over-predicts the comoving density of low-mass galaxies. 
We find that in the low and intermediate mass range, $10^{8.5}\ge M_{\star}\le 10^{11.5}\Msun$, the co-moving density functions predicted by our model show a double power-law shape, with a steeper slope at the low mass end of the model distribution. At the highest redshifts, $z \simeq 6.0$, the density function keeps this shape even for the most massive model galaxies, $M_{\star}\ge 10^{11.5}\Msun$. This power-law shape is the signature of the continuous competition between the disruption and the progressive fragmentation of the gas through the turbulent cascade. Some galaxies, those with stellar masses above $10^{10.5}~\Msun$, have already substantially formed at $z \simeq 6.0$. The turbulent cascade regulates the star formation in intermediate- and low-mass objects and keeps a sufficient amount of diffuse and fragmented non star-forming gas that actively feeds star formation until the formation of the most massive observed galaxies. 

\begin{figure*}[t!]
  \begin{center}
  \includegraphics[width=\linewidth]{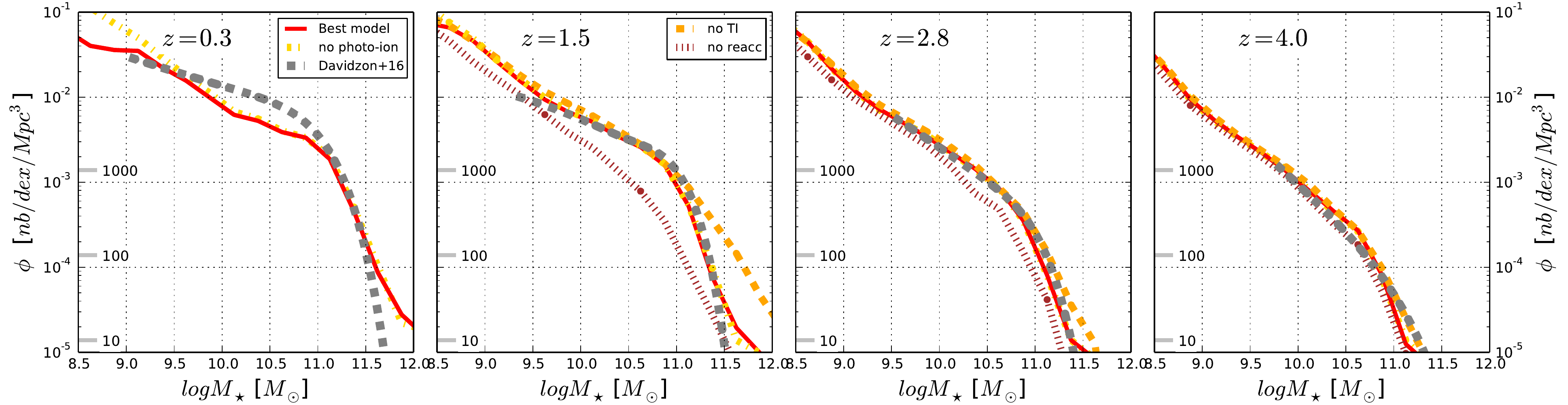}
  \caption{The impact of our main prescriptions for regulating star formation and gas accretion, photo-ionisation, re-accretion and thermal instabilities on the stellar mass assembly of model galaxies (i.e., the co-moving density of model galaxies as a function of redshift). The solid red line shows our fiducial model using the full suite of prescriptions presented here. The dot-dashed gold lines indicate our model without photo-ionisation. In the three right-most panels, the dotted brown lines and the dashed orange lines are predictions of the stellar mass functions without including the prescriptions for re-accretion of gas over two halo crossing times and thermal instabilities, respectively (as indicated in the legends of the two left-most panels). These are compared to the observational estimates of the co-moving density of galaxies from \cite{Davidzon_2016}. The four different panels are for four different redshift bins, $z$ = 0.3, 1.5, 2.8, and 4 (from the left-most to the right-most panel respectively).}
  \label{fig:stellar_mass_function_comparison}
  \end{center}
\end{figure*}

As the redshift decreases, the power law-shape is broken at the high mass end of the distribution and an exponentially declining function is progressively formed. In agreement with observational estimates, the knee of our predicted co-moving number density functions develops around $10^{10.75}\Msun$ at $z\simeq5.0$ and progressively shifts to $10^{11.0}\Msun$ by $z\simeq0.1$. The break appears when the star formation is strongly reduced or quenched, i.e., when all the fresh gas contained in the disk is consumed. At the high mass end of the galaxy distribution, gas is accreted via the hot mode (Sect.~\ref{sec:gas_accretion} and Fig.~\ref{fig:acc_rate}). In, for example, \cite{Guo_2011}, \cite{Croton_2006} or \cite{Somerville_2008}, the reduction of the gas accretion is due to the power of the AGN heating the halo gas and halting accretion. To obtain the necessary heating, such prescriptions need an increasing and constant production of power from AGN. In \GAS, much of the reduction or halting of gas accretion is a result of the increase in the thermally unstable hot gas phase surrounding galaxies. As shown in Fig. \ref{fig:stellar_mass_function_comparison}, the impact of thermal instabilities appears to be progressive and its impact starts to become significant at $z\sim4.0$ for the most massive galaxies. The effect is clearly visible for the galaxy mass function at $z = 2.1$. Without such regulation, the continuous accretion of gas leads to the formation of very massive, $10^{12.0}\Msun$, galaxies at $z\simeq1.5$. At very high masses, the reduction of the gas accretion onto galaxies is not sufficient to quench completely the star formation. Indeed, some galaxies have stellar masses more than $10^{11.5}\Msun$ are predicted in our model but are not observed \citep[e.g.,][]{Davidzon_2016, Ilbert_2013, Yang_2009, Baldry_2008}. This is due to a progressive decrease of the volume filling factor of thermal instabilities in the most massive galaxies (Fig.\ref{fig:TI_VFF}). A slight increase of the thermal instabilities propagation efficiency, $\varepsilon_{TI}$, could limit the growth of these very massive model galaxies. 

Compared to the \textit{ad-hoc} recipe ($\propto M_{h}^3$) presented previously in \citet{Cousin_2015a}, the new regulation prescription based on the turbulent cascade is in better agreement with the comoving densities of low-mass galaxies. The \textit{ad-hoc} recipe was clearly not appropriate for capturing the necessary ingredients to produce sufficient numbers of low mass galaxies and especially those at high redshifts (Fig.\ref{fig:stellar_mass_function_comparison_other_SAM}). On the contrary, the stellar mass functions predicted by \cite{Guo_2011} clearly over-produce the comoving density of low/intermediate galaxies, $M_{\star}<10^{10.5}\Msun$. Their  prescriptions for regulating star formation do not have the correct dependencies as a function of the stellar mass. Gas is converted to stars too efficiently in low/intermediate mass galaxies, hence producing an excess. Furthermore, a lack of gas in more massive galaxies means that their models under-estimate the number of massive galaxies. 

In fact, to elucidate the role played by various physical prescriptions we now include in \GAS, we compare four different
model configurations: our best complete model and three alternative versions in which: i) the gas re-accretion prescription is disabled; ii) thermal instabilities are not considered; and, iii) the photo-ionisation prescription is switched off. These three alternative versions allow us to determine the effective impact of those prescriptions in regulating the mass growth of the ensemble of model galaxies (Fig.~\ref{fig:stellar_mass_function_comparison}). At the highest redshift for which we made this comparison, $z$=4, we see that turning off all 3 of these prescriptions has little impact on the modelled stellar mass distribution function. As we decrease in redshift, we see progressively greater impact for these 3 processes on the stellar mass distribution. The lack of thermal instabilities in the hot gas in the halo, results in overly massive galaxies by z$\approx$2-3 but leaves the number of lower mass galaxies, those with M$_{\star}$$\la$10$^{11}~\Msun$ essentially unchanged.  Photo-ionisation appears to only impact the co-moving number density of galaxies at low redshift and for galaxies with low stellar masses. At $z<1$, the power-law shape in the co-moving density of low mass galaxies, i.e., those with $M_{\star}<10^{8.5}\Msun$, is broken and becomes flatter as the redshift decreases. This trend is a result of a progressive reduction in the effective gas accretion with decreasing redshift and is a direct consequence of the photo-ionisation prescription (Sect. \ref{sec:gas_accretion}, Fig.~\ref{fig:normalized_baryonic_fraction}).
The process with the largest impact on the co-moving number density of galaxies appears to be the re-accretion of gas (Sect.~\ref{sec:2-orbits_re-accretion}). Its impact on the galaxy co-moving density distribution starts to become evident at $z$$\approx$3 (Fig.~\ref{fig:stellar_mass_function_comparison}). Without re-accretion, the model strongly under-predicts the comoving density of almost all stellar masses but especially so for intermediate-mass galaxies. 

\subsection{Evolution of disk properties}
\label{sec:main_disk_properties}

\begin{figure*}[t!]
  \begin{center}
    \includegraphics[width=\linewidth]{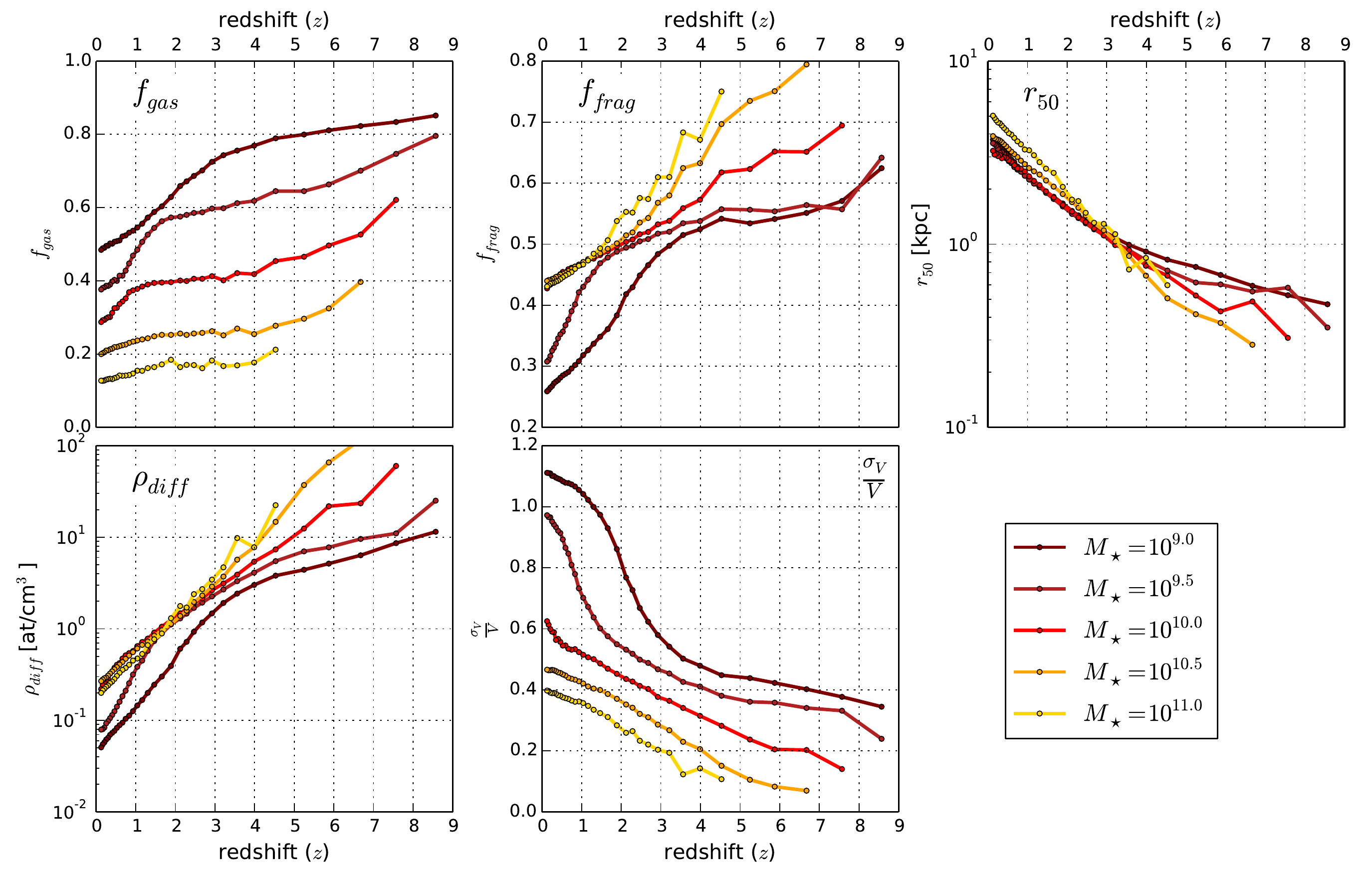}
  \caption{The evolution with redshift of five main disk properties of disk galaxies which lie along the star-forming galaxy main sequence shown for a range of mass bins (see legend at the bottom right panel for the stellar mass values for each curve in the panels). Upper left: The gas mass fraction, $f_{gas}$, as a function of redshift for 5 stellar mass bins. Upper centre: The fraction of the gas which is fragmented, $f_{frag}$. Upper right: The average half mass radius, $r_{50}$, of the ensemble of star-forming model galaxies. Lower left: The diffuse gas density, $\rho_{diff}$. Lower right: The ratio of the gas velocity dispersion to orbital velocity, $\frac{\sigma_V}{V}$.}
  \label{fig:main_disk_properties}
  \end{center}
\end{figure*}

We discuss with Fig. \ref{fig:main_disk_properties} the evolution with both the stellar mass and the redshift of five main properties of star-forming galaxies. We present the evolution of: 
\begin{itemize}

\item {The gas mass fraction, $f_{gas} = \frac{M_{gas}}{M_{gas}+M_{\star}}$. $M_{gas}$ is the total mass of the three gas reservoirs combined for a given disk galaxy.}  

\item {The fragmented gas fraction, $f_{frag} = \frac{M_{frag} + M_{sfg}}{M_{frag} + M_{sfg} + M_{diff}}$.}

\item {The galaxy half mass radius, $r_{50}$.}

\item {The diffuse gas density, $\rho_{diff}$. This density is computed based on the mass of diffuse gas, a disk with an external radius of $22\times r_d$ and a scale height, $h_d$. We also assume a mean molecular mass, $\mu$= 0.62.}

\item {The ratio of the gas velocity dispersion to orbital velocity, $\frac{\sigma_v}{V}$. The dispersion velocity is that of the diffuse gas. The orbital velocity is computed at 2.2$r_d$ \citep{Pelliccia_2017}.}

\end{itemize}

The following analysis is based on model star-forming galaxies that have been extracted at a variety of redshifts. We define star forming as those galaxies which lie above 25\% of the mean of the relation between the star-formation rate and stellar mass \citep[the ``main sequence'' of star-forming galaxies][their Eq.~9]{Schreiber_2015}. Within this sample of model galaxies, we calculate median of the ensemble binned in redshift and stellar mass for each of the quantities listed above.
We first focus on the gas mass fraction, $f_{gas}$. At a given stellar mass, the highest redshift galaxies have the highest gas fractions while galaxies with larger stellar masses have lower gas fractions. If we consider both the increase in mass with decreasing redshift, these tracks implies that the gas mass fraction globally decreases as galaxies grow and evolve. At low redshift, $z < 0.5$, the gas mass fractions of star-forming galaxies lie between 15\%-50\% depending on the stellar mass. In two samples of star-forming galaxies, evolving around $z = 1.5$, \cite{Daddi_2010} measured a gas mass fraction higher than our predictions. At $z = 1.5$, for an average stellar mass of $M_{\star} = 10^{10.5}\Msun$, the gas mass fraction is distributed as [p15, p50, p85] = [0.18, 0.25, 0.36] in comparison to $f_{gas}\simeq$0.6 measured by \cite{Daddi_2010}. In the same redshift slice, for an average stellar mass of $M_{\star} = 10^{11}\Msun$ we measured [p15, p50, p85] = [0.11, 0.16, 0.26] in comparison to $f_{gas}\simeq$0.5 measured by \cite{Daddi_2010}.

In parallel to the progressive decrease of the gas-mass fraction, the distribution of this gas between the diffuse and the fragmented/dense gas also evolves. We first note that at a given stellar mass, the fragmented gas mass fraction is an increasing function of the redshift. Galaxies formed at higher redshift contain a larger fraction of fragmented gas than galaxies evolving at lower redshift with a similar mass. At high redshift $z > 2.5$, the fragmented gas mass fraction is a clear decreasing function of the stellar mass. The hierarchy in mass is less clear at lower redshifts. We note that galaxies hosting a stellar mass larger than $10^{10}\Msun$ stabilise their fragmented gas mass fraction at around 42\%. For less massive galaxies, $<10^{10}\Msun$, the fragmented gas fraction strongly decreases and reaches low values distributed between 25\% and 33\% at $z = 0.1$.

\begin{figure*}[t!]
  \begin{center}
    \includegraphics[width=\linewidth]{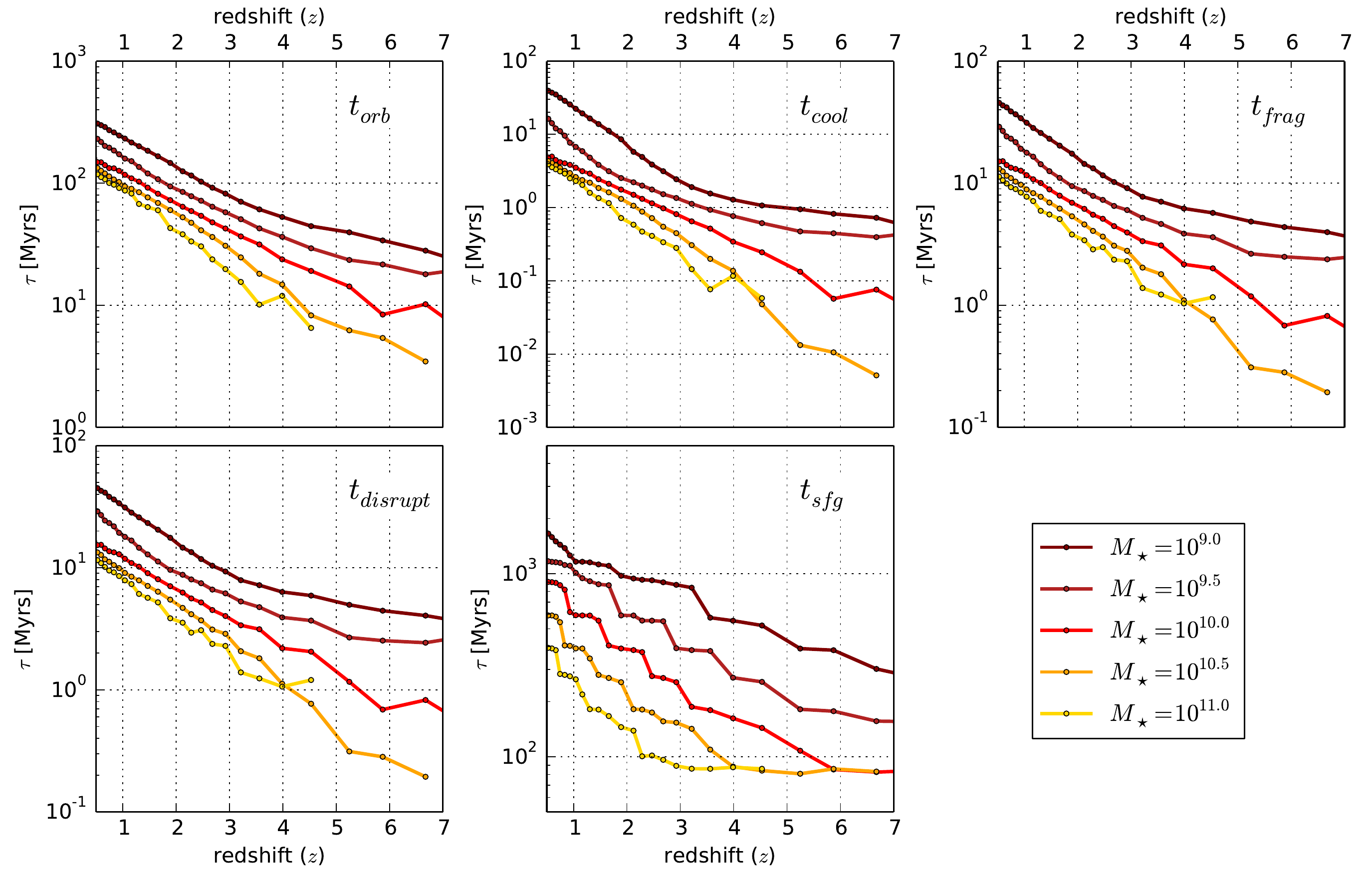}
  \caption{The evolution with redshift of five main timescales for regulating the growth of disk galaxies which lie along the star-forming galaxy main sequence shown for a range of redshift bins (see legend in the bottom right panel for the stellar mass value for each curve in the panels). Upper left: The orbital timescale, $t_{orb}$. Upper centre: The characteristic cooling timescale, $t_{cool}$. Upper right: The fragmentation timescale of the gas, $t_{frag}$. Lower left: The gas disruption timescale, $t_{disrupt}$. Bottom left:  The gas depletion timescale, $t_{sfg}$.}
  \label{fig:main_disk_time scales}
  \end{center}
\end{figure*}

The gas content clearly evolves with time: gas rich and strongly fragmented galaxies evolve through structures dominated by stars and in which the gas is mainly diffuse. This evolution in gaseous and stellar contents is following the evolution of galaxy size: At a given stellar mass, the disc half mass radius is a decreasing function of the redshift. At redshift $z>3.0$, the disc half mass radius is a decreasing function of the stellar mass. Massive galaxies appear more compact than low-massive galaxies. For all stellar mass bin analysed, the average half mass radius is strictly lower than 1~kpc. At $z<3.0$, we note the opposite trend: the average half mass radius is strictly larger than 1~kpc and is an increasing function of the stellar mass.

The anti-correlated evolution of the gas content and disk size directly impacts the density of the diffuse gas which is the main driver of the GMC feeding rate (Sect.~\ref{sec:diffuse_gas_reservoir}, Eq.~\ref{eq:fragmenting_rate}). At a given stellar mass, the diffuse gas density appears to be an increasing function of the redshift. In galaxies formed at higher redshift, the gas is denser than in galaxies of a same stellar mass formed at lower redshift. At a $z>3$, the diffuse gas density is an increasing function of the stellar mass. At $z<3$, the most massive galaxies see their diffuse gas density converge to 0.2-0.3 atoms/cm$^3$. The less massive galaxies ($\le10^{10}\Msun$) see their diffuse gas density strongly decrease and reach a value of $<0.1$atoms/cm$^3$. This evolution is mainly linked to the decrease of the fragmented gas-mass fraction discussed previously. 

This decrease also impacts the dispersion to orbital velocities ratio, $\frac{\sigma_v}{V}$. This ratio appears to be a decreasing function of both the redshift and the stellar mass. At $z\le 3.0$, we measure a strong increase of $\frac{\sigma_v}{V}$ in the two lowest stellar mass bins. This strong evolution observed at low redshift is linked to the injection of kinetic energy by the latest-formed SN populations in a gas that becomes rare and diffuse. We recall that the velocity dispersion is computed at the disk scale height and takes into account only the diffuse gas mass. Less gas means a turbulent energy budget (Sect. \ref{sec:Disruption_rate}) divided into a smaller number of gas atoms and therefore a higher average velocity per mass unit. In parallel, the decrease with the redshift of the ratio $\frac{\sigma_v}{V}$ is also due to the increase with the redshift of the orbital velocity traced by the decrease of the average disk orbital timescale (see upper left panel of Fig. \ref{fig:main_disk_time scales}). By tracing average evolutions from low mass at high redshift to high mass at low redshift, the average galaxy trends indicate that the $\frac{\sigma_v}{V}$ ratio slightly decreases through the evolution.   

\subsection{Timescales}
\label{sec:time scales}

\begin{table*}[th!]
  \begin{center}
      \begin{tiny}
      \begin{tabular}{l|ccccc}
        \hline
                & $10^{9.0}\pm 0.2\Msun$ & $10^{10.0}\pm 0.2\Msun$ & $10^{10.25}\pm 0.2\Msun$ & $10^{10.75}\pm 0.2\Msun$ & $10^{11.25}\pm 0.2\Msun$ \\[1.0ex]
        \hline
\textbf{Full Sample} & & & & & \\
\hline
259090 & 11.1\% & 2.0\% & 1.4\% & 0.6\% & 0.0\% \\[1.0ex]
\hline
\textbf{timescales} & [p15, p50, p85] & [p15, p50, p85] & [p15, p50, p85] & [p15, p50, p85] & [p15, p50, p85] \\[1.0ex]
\hline
$t_{orb}$  [Myr]           & [71.48, 132.42, 210.06] & [32.38, 67.23, 118.72] & [29.12, 61.81, 108.64] & [17.38, 45.52, 100.14] & [16.63, 36.94, 106.27] \\[1.0ex]
$t_{cool}$ [Myr]           & [2.281, 7.913, 28.065] & [0.616, 1.791, 5.333] & [0.527, 1.527, 4.041] & [0.263, 1.140, 4.030] & [0.326, 1.325, 5.042] \\[1.0ex]
$t_{frag}$ [Myr]            & [7.929, 16.453, 31.759] & [3.152, 6.811, 13.585] & [2.684, 5.865, 11.290] & [1.587, 4.311, 9.970] & [1.790, 3.883, 11.050] \\[1.0ex]
$t_{disrupt}$ [Myr]        & [7.732, 15.892, 28.138] & [3.111, 6.672, 12.628] & [2.642, 5.874, 10.966] & [1.481, 4.328, 9.919] & [1.661, 3.733, 10.815] \\[1.0ex]
$t_{sfg}$ [Myr]            & [864.5, 974.5, 1577.5] & [255.8, 390.2, 903.2] & [181.0, 280.2, 615.8] & [139.5, 182.3, 878.4] & [91.7, 181.1, 897.2] \\[1.0ex]
\hline
\textbf{ratios} & [p15,p50,p85] & [p15,p50,p85] & [p15,p50,p85] & [p15,p50,p85] & [p15,p50,p85] \\[1.0ex]
\hline
$t_{cool}/t_{orb}$ & [0.029, 0.055, 0.143] & [0.018, 0.025, 0.042] & [0.017, 0.023, 0.035] & [0.012, 0.022, 0.042] & [0.014, 0.029, 0.058] \\[1.0ex]
$\mathcal{T}_{cool}/t_{cool}$ & [1.781, 2.694, 3.556] & [2.982, 3.488, 3.746] & [3.171, 3.472, 3.757] & [3.058, 3.445, 4.199] & [2.963, 3.312, 4.314] \\[1.0ex]
$t_{frag}/t_{orb}$ & [0.105, 0.118, 0.151] & [0.088, 0.096, 0.115] & [0.083, 0.092, 0.107] & [0.075, 0.088, 0.113] & [0.075, 0.093, 0.124] \\[1.0ex]
$t_{sfg}/t_{orb}$ & [5.48, 8.30, 18.55] & [4.13, 6.55, 13.47] & [3.36, 5.36, 11.69] & [2.57, 5.36, 18.16] & [2.12, 6.16, 32.75] \\[1.0ex]
$t_{sfg}/t_{frag}$ & [43.63, 69.07, 137.39] & [41.00, 65.86, 123.26] & [35.63, 57.21, 115.10] & [29.59, 57.33, 169.20] & [24.12, 60.52, 235.34] \\[1.0ex]
\hline
 & & & & & \\
\hline
                    & $10^{9.0}\pm 0.2\Msun$ & $10^{10.0}\pm 0.2\Msun$ & $10^{10.25}\pm 0.2\Msun$ & $10^{10.75}\pm 0.2\Msun$ & $10^{11.25}\pm 0.2\Msun$ \\[1.0ex]
\hline
\textbf{Star Forming Sample} & & & & & \\
\hline
76762 & 2.2\% & 1.5\% & 1.1\% & 0.3\% & 0.0\% \\[1.0ex]
\hline
\textbf{timescales} & [p15, p50, p85] & [p15, p50, p85] & [p15, p50, p85] & [p15, p50, p85] & [p15, p50, p85] \\[1.0ex]
\hline
$t_{orb}$  [Myr]           & [65.87, 129.84, 209.79] & [30.95, 65.21, 115.34] & [28.80, 60.72, 105.58] & [18.18, 45.44, 98.73] & [11.16, 47.99, 109.13] \\[1.0ex]
$t_{cool}$ [Myr]           & [1.738, 6.264, 16.940] & [0.547, 1.504, 3.506] & [0.471, 1.355, 2.933] & [0.191, 0.804, 2.395] & [0.070, 0.801, 3.186] \\[1.0ex]
$\mathcal{T}_{cool}$ [Myr] & [6.024, 17.445, 36.808] & [1.967, 5.358, 11.648] & [1.673, 4.771, 9.871] & [0.794, 2.929, 8.081] & [0.355, 3.290, 9.476] \\
$t_{frag}$ [Myr]            & [7.059, 14.864, 26.348] & [2.914, 6.132, 11.352] & [2.542, 5.504, 9.915] & [1.407, 3.806, 8.482] & [0.713, 3.976, 9.133] \\[1.0ex]
$t_{disrupt}$ [Myr]        & [7.166, 15.119, 26.448] & [2.937, 6.252, 11.630] & [2.584, 5.608, 10.130] & [1.429, 3.906, 8.714] & [0.714, 4.146, 9.505] \\[1.0ex]
$t_{sfg}$ [Myr]            & [834.8, 943.1, 1162.4] & [255.4, 382.5, 590.6] & [174.4, 269.8, 400.7] & [105.2, 166.7, 262.0] & [69.4, 99.3, 175.9] \\[1.0ex]
\hline
\textbf{ratios} & [p15,p50,p85] & [p15,p50,p85] & [p15,p50,p85] & [p15,p50,p85] & [p15,p50,p85] \\[1.0ex]
\hline
$t_{cool}/t_{orb}$ & [0.029, 0.055, 0.143] & [0.018, 0.025, 0.042] & [0.017, 0.023, 0.035] & [0.012, 0.022, 0.042] & [0.014, 0.029, 0.058] \\[1.0ex]
$\mathcal{T}_{cool}/t_{cool}$ & [1.781, 2.694, 3.556] & [2.982, 3.488, 3.746] & [3.171, 3.472, 3.757] & [3.058, 3.445, 4.199] & [2.963, 3.312, 4.314] \\[1.0ex]
$t_{frag}/t_{orb}$ & [0.104, 0.115, 0.132] & [0.087, 0.094, 0.104] & [0.082, 0.091, 0.099] & [0.072, 0.083, 0.092] & [0.067, 0.081, 0.095] \\[1.0ex]
$t_{sfg}/t_{orb}$ & [5.33, 7.61, 12.45] & [3.88, 5.81, 9.11] & [3.17, 4.76, 7.60] & [2.07, 3.70, 7.31] & [1.32, 2.30, 7.58] \\[1.0ex]
$t_{sfg}/t_{frag}$ & [43.41, 64.93, 114.44] & [39.95, 60.73, 97.30] & [34.20, 51.90, 85.50] & [24.63, 42.88, 89.36] & [14.79, 30.06, 124.96] \\[1.0ex]
\hline
        \end{tabular}
      \end{tiny}
  \end{center}
  \caption{\footnotesize{Statistics of the main timescales in the gas cycle. The model galaxies used for this analysis all lie at $z=2.1$. Median values and 15\% and 85\% percentiles are given for five different stellar mass bins. The upper table shows statistics of the fully resolved galaxy sample. The lower table shows the star-forming sample. In each sub-Table, the upper part shows absolute values of the timescales and the lower part shows the ratios with the orbital time or gas fragmentation timescales.}}
  \label{tab:main_disk_time scales}
\end{table*}

The gas cycle presented in Sect.~\ref{sec:Gas_cycle} is based on three gas reservoirs and different exchange rates between these reservoirs. From these transfer rates and knowing the mass in each of these gas reservoirs, we can deduce the various timescales. Evolution with both the redshift and the stellar mass of five main timescales are shown in Fig.~\ref{fig:main_disk_time scales}. We focus on:
\begin{itemize}

	\item the orbital time $t_{orb}$ estimated at a radius, $2.2r_d$ \citep[e.g.,][]{Pelliccia_2017};
  
	\item the cooling time-scale, $t_{cool}$, given by Eq. \ref{eq:cooling_time_function};
 
    \item the fragmentation time-scale,  $t_{frag}=\frac{M_{frag}+M_{sfg}}{\dot{M}_{frag}}$, defined as the time required to double the fragmented gas mass;
   
    \item the disruption timescale, $t_{disrupt}=\frac{M_{frag}+M_{sfg}}{\dot{M}_{disrupt}}$, defined as the time required to deplete the fragmented gas via the kinetic energy provided by SN and/or AGN;
    
    \item the gas depletion time of the fragmented gas, $t_{sfg}=\frac{M_{frag}}{\dot{M}_{sfg}}$, defined as the time required to convert all the fragmented gas and star-forming gas into stars. As the star-formation timescale is less than all other timescales, $t_{sfg}$, is essentially the time required to convert all of the fragmented gas into stars.
    
\end{itemize} 

We present details of the model output by providing the medians and the 15\% and 85\% percentiles of these characteristic timescales (Table~\ref{tab:main_disk_time scales}). We compile the statistics for both the full and star-forming samples of model galaxies at $z = 2.1$. These estimates given in four different stellar mass bins, 10$^{9}$, 10$^{10}$, 10$^{10.5}$, and 10$^{11.25}$ $\Msun$. Even if the median trends presented in Fig.~\ref{fig:main_disk_time scales} suggest some regularity in galaxy properties as a function of mass, all timescales at constant mass have significant scatter (see Tab.~\ref{tab:main_disk_time scales}). This large scatter indicates that the dynamics for any one galaxy or ensemble of galaxies is not regular or simple, but is a complex interaction of the various processes included in the model.

First, we first focus on the disk orbital timescale which plays a key role in controlling processes that occur over long timescales. Depending on redshift and stellar mass, the orbital timescale ranges between $\simeq$4 Myrs and $\simeq$300 Myrs. At fixed stellar mass, the orbital timescale increases with the redshift. At fixed redshift, it is a decreasing function of the stellar mass. The range of orbital timescales calculated using the \GAS\, model are in good agreement with orbital timescales estimated for local galaxies \citep[e.g.,][]{Kennicutt_1998, Leroy_2008, Colombo_2018}. 

The fragmentation of the ISM is driven by energy injection on scales of and larger than the disc height via the condensation of the diffuse gas. The rate of fragmentation depends on the radiative cooling rate of the diffuse warm gas. The characteristic cooling time is dependent on redshift and galaxy mass and ranges over five orders-of-magnitude (Eq.~\ref{eq:cooling_time_function}). At a given stellar mass, the characteristic cooling timescale is a decreasing function of the redshift (Fig.~\ref{fig:main_disk_time scales}). This trend is driven by the increase with the redshift of the diffuse gas density. At a given redshift, the characteristic cooling time is a decreasing function of the mass. These two trends together means that the characteristic cooling timescale of a median galaxy track decreases with both redshift and stellar mass (Fig.~\ref{fig:main_disk_time scales}).
At $z = 2.1$, the characteristic cooling timescale of our star-forming galaxies sample is distributed between $\simeq$0.8 Myr and $\simeq$6.2 Myr. These values are slightly smaller than in the fully resolved galaxy sample of $\simeq$1.3 Myr and $\simeq$8.0 Myr (Table~\ref{tab:main_disk_time scales}).
At all redshifts and for all stellar masses, the characteristic cooling timescale is significantly shorter than the orbital timescale ($<$6\%). If no additional mass were added to the reservoir of diffuse gas, its condensation, which is mainly governed by the characteristic cooling timescale, would deplete completely in less than an orbital time.
The short cooling and condensation timescales helps to drive the overall complex dynamics
of the gas cycle in model galaxies as alluded to earlier. 

To provide an indication of how much of the diffuse gas is condensing at any given time, we tabulate the mass fraction of the diffuse gas (Table~\ref{tab:main_disk_time scales}). The mass fraction of the diffuse gas that is condensing can be estimated by comparing the effective cooling time (Eq.~\ref{eq:cooling_clock}) to the characteristic cooling timescale (Eq.~\ref{eq:cooling_time_function}). At $z=2.1$, for the five stellar mass bins tabulated, the median values of the ratio, $\mathcal{T}_{cool}/t_{cool}$, range between $\simeq$2.7 and $\simeq$3.5 for both model galaxy samples. Low mass galaxies have the lowest ratios. The ratio is approximately constant, $\simeq 3.5$, for galaxies with $M_{\star}\ge 10^{10}\Msun$.

The fragmentation and disruption timescales have very similar dependencies on redshift and stellar mass (Fig.~\ref{fig:main_disk_time scales}). These two timescales show a decreasing trend with both redshift and stellar mass. At $z = 2.1$, for the star-forming galaxy sample, the fragmenting timescale and the disrupting timescale are distributed between 4 and 15 Myrs. Those values correspond to $\simeq$10\% of the disk orbital timescale. As for the cooling timescale, the fragmentation and the disruption timescales of our star-forming galaxy sample are marginally smaller than those of the full galaxy sample. The disruption timescale is always (slightly) larger than the fragmentation timescale (Table~\ref{tab:main_disk_time scales}) and thus gas fragmentation is always faster than gas disruption in \GAS. 
The short fragmentation timescales indicate that the growth of GMCs is a continuous and efficient process. Timescales measured in our model are fully compatible with GMC gas accretion rates found via numerical simulations \cite[e.g.][]{Vazquez_2010, Colin_2013}.

Working against the gas cooling and fragmentation is the injection of energy by SNs and AGN. This energy injection is used in \GAS\ to disrupt the gas and transfer some of the fragmented gas back into the reservoir of diffuse gas. Hydrodynamical simulations find that disruption timescales of about 10--20 Myr \citep{Colin_2013,Dobbs_2015}. This range of values is fully consistent with our predictions. 

\subsubsection{Impact of fragmentation and disruption on the life cycle of GMCs}

The small differences between the condensation and disruption timescales implies that the  the total mass of fragmented gas in GMCs stays approximately constant and that significant fraction of the fragmented mass is continuously regenerated. From this equilibrium, there are two evolutionary paths for GMCs that depend on both the scale and mass of GMCs.

\begin{itemize}

\item{Low mass GMCs, those formed in galaxies with short or intermediate disk-scale heights, $h_d\le$50pc, the mass disrupted by SN kinetic energy injection is close to the total of low mass GMCs. In such circumstances, after only few SN cycles, GMCs are fully disrupted. These disrupted GMCs are constantly replenished through the formation of new GMCs through the condensation of diffuse gas. Thus the disruption/fragmentation timescales are approximately the lifetime of the GMCs in our model. Our results are in reasonable agreement with the sizes and short GMC lifetimes estimated in local galaxies, 25-70~pc and 17$\pm$4~Myr \citep{Murray_2011,Miura_2012,Meidt_2015}}.

\item{More massive GMCs resist being disrupted by SN. It is the competition between accretion and disruption which regulates the mass of GMCs \citep{Vazquez_2010}. Due to this competition, GMCs have to be therefore considered as dynamical structures where the gas is just \textit{passing through} from the diffuse-gas state to the star-forming gas state.}

\end{itemize}

SN/AGN-driven gas disruption slows significantly the rate at which gas fragments. In small GMCs, the cloud structure and therefore the sites of star formation can be completely destroyed after only a few SN cycles. In more massive GMCs, the fragmentation is also significantly slowed. Indeed, gas recently added to the reservoir of diffuse gas starts to fragment on larger scale than the gas within the GMCs. In our prescription, this constant renewal and exchange of the gas in these reservoirs and the impact of the added gas to the overall cascade of fragmenting gas is fully accounted for via the mass-weighted fragmentation clock (Eq.~\ref{eq:t_str}).

\subsubsection{Gas depletion timescale}

Depending on the redshift and the stellar mass of a model galaxy, the gas depletion timescale, $t_{sfg}$, ranges from 80 Myr to 2~Gyr (Fig.~\ref{fig:main_disk_time scales}). The upper end of this range is consistent with with gas depletion timescale measured in local spiral galaxies \citep[e.g.,][]{Leroy_2013, Colombo_2018} and the values predicted at high redshift, z$\approx$2-3, are also generally consistent with the values estimated for distant star-forming galaxies \citep[e.g.,][]{Daddi_2010, Genzel_2010}. Gas depletion timescales are a decreasing function of both the redshift and the stellar mass. Above $M_{\star} \ge 10^{10.5}\Msun$ and at z > 3.0, the average gas depletion timescale reaches a lower-limit close to 80~Myr. 

The $t_{sfg}/t_{frag}$ ratio provides an estimate of the efficiency of star formation in GMCs. The number of condensation/disruption-cycles and the depletion timescale of fragmented gas imply that faster cycles lead to shorter depletion timescales. At z = 2.1, median values extracted from our star-forming sample, indicate that between $\simeq$30 and $\simeq$65 cycles are required to convert fragmented gas into star-forming gas (and therefore into stars). This number of gas cycles is consistent with gas dynamics and number of cycles estimated in hydrodynamic simulations \citep[e.g.,][]{Semenov_2017}. In the star-forming sample, the average number of condensation/disruption-cycles is strictly a decreasing function of the stellar mass. However, in the full galaxy sample, a minimum is reach for models galaxies with mass of $10^{10.25}\Msun$.  The fragmenting gas in galaxies less and more massive than $10^{10.25}\Msun$ need more cycles to convert the fragmented gas into stars. During each cycle only a small fraction, 1\%-5\%, of the fragmented gas stops fragmenting and is available for the star formation. In fact, the majority of the gas that forms GMCs is recycled in of-order one to a few dynamical times or simply remains in a state that cannot form stars. We obtain star-formation efficiencies between $\simeq$1 - 5\%. Such values are consistent with observational estimates \citep[e.g.,][]{Wong_2002, Leroy_2008, Murray_2010, Andre_2013a, Heiderman_2010} and efficiencies measured in hydrodynamic simulations \citep[e.g.,][]{Semenov_2017, Kimm_2017}. 

In summary, the gas cycle implemented in the new \GAS\ model attempts to capture the complex dynamics of the gas in galaxies.  In our prescriptions, gas is continuously exchanged between the diffuse and the fragmented non-star forming phases. A large number of condensation and disruption cycles, 30-70, are needed to progressively convert diffuse gas into very dense star-forming gas and then stars. The characteristic timescales of $\simeq$15~Myrs of these exchange rates is only a small fraction of the orbital time-scale. Under such conditions, the fragmented gas is converted into star-forming gas and stars continuously and the gas spends the majority of its time in a non-star-forming gas phase. 

\section{Discussion and Conclusions}
\label{sec:discussion}

We present a new semi-analytical model, \GAS, in which we implemented a more realistic gas cycle than has been previously implemented in a semi-analytical model. We introduced a prescription for delaying star formation which is underpinned by progressively fragmenting the gas. The formation of giant molecular clouds, filaments and cold cores takes time and therefore at a given instant only a small fraction of the total gas is in the form of pre-stellar cores. The majority of the gas is not in a form that is immediately available to form stars. Within this framework, we account for the continuous dissipation of the turbulent kinetic energy through the different ISM scales and phases. We implemented an approximate multi-phase ISM where the gas cycles between a warm diffuse phase, a cold fragmented non-star-forming phase, and a very dense star-forming gas phase. 
Diffuse warm gas is progressively fragmented following the effective cooling time. Even if the overall depletion timescale of the fragmented gas is, on average, proportional to the disk orbital time-scale, we measure a large scatter in both the depletion time-scale and the disk orbital time-scale. This proves that prescriptions that rely solely on the disk orbital timescale do not capture properly the complexity of gas cycles in galaxies. Smaller characteristic timescales, such as the  cooling and/or energy injection timescales due to supernovae and AGN are also important. This energy input on shorter timescales,
efficiently disrupts the star forming gas. The large-scale velocity dispersion of the diffuse gas is also maintained by SNs/AGN kinetic energy injection. A fraction of the gas can also be ejected from disks by SN explosions but the characteristic time-scale of ejection is in average 10 times larger than the fragmenting/disruption cycle time-scale. While the SN/AGN kinetic energy injection in the ISM regulates the star-formation rate in low mass galaxies, our model galaxies retain a large fraction of the gas. This ``local'' gas is then used with a progressively increasing efficiency necessary to build massive, $M_{star} > 10^{11}\Msun$, galaxies in the early  Universe, $z\in[4$; $6]$. 

Star formation occurs only in the very dense gas, which in our model is a product of continuous fragmentation over a wide range of scales. This new complete gas cycle strongly regulates the  star formation in our modelled galaxies. Only a few percent of the available fragmented gas is converted into stars during a GMC life cycle. By taking into account the fragmented gas on a galaxy scale, our model is able to reproduce the standard Schmidt-Kennicutt law. Our estimated gas depletion timescales, which are directly related to the time it takes for gas to fragment, are in good agreement with observational estimates. 

This new gas regulation cycle leads to very good agreement with the observed stellar mass function over a wide range of redshifts, $z\in[0.8$; $6]$. The ability of our model to catch the stellar mass assembly in high redshift galaxies as been already used with success in \cite{Lagache_2018}. Galaxy properties (gas/stars contents, metallicities and FUV fluxes ...) predicted by our model have been post-processed to successfully predict the [CII] luminosity functions at high redshifts and allowed us to explore the main characteristics of this emission.

At $z<0.8$, we find some discrepancies between the predicted and the observed stellar mass functions. Specifically, an over-density of low-mass galaxies and the under-density of intermediate mass galaxies modelled at z<0.8 could probably be solved by slightly increasing the efficiency of photo-ionization in our prescription.

At $z<4.0$, to reduce the efficiency of or to stop the growth of stellar mass in massive galaxies, the accretion of gas and its transformation into stars has to be reduced or stopped. Previous semi-analytical models implemented strong AGN feedback to quench gas accretion onto massive galaxies. In those models, a significant fraction of the power produced by the AGN is directly used to reduce the cooling of the hot halo gas. This implies a constant AGN power production, which may not be consistent with our understanding of AGN variability \citep[see, e.g.,][and references therein]{hickox2014,stanley2015,volonteri2015}. For massive galaxies, we propose an alternative model which regulates the radiative cooling and gas accretion onto galaxies. In parallel to radiative cooling, we implemented the development of thermal instabilities creating warm gas surrounding the galaxy. The growth of these thermal instabilities in the central region of the halo progressively reduces or halts the cooling and dissipation of kinetic energy in the halo and therefore reduces or stops the accretion onto galaxies. This process of regulation is a natural outcome of the growth of thermal instabilities in the hot halo gas and does not depend on the power produced by AGN. 

However, our actual efficiency parameters do not fully stop gas accretion onto very massive galaxies. In some massive dark-matter halos, radiative cooling restarts (VFF $<$ 1.0) even if its hot gaseous halo has been fully quenched previously (VFF = 1.0). The recovery of the gas accretion leads to the formation of some ($<$10) unobserved overly massive galaxies at z$\simeq$0.3. Our prescription of thermal instability growth in the hot gas phase is governed by a set of two parameters which can be further refined to overcome this problem. 

In our model of the gas cycle, we assume that the gas initially fragments at the disk scale height with the progressive and continuous formation of over-densities which are akin to observed GMCs. The fragmentation of the gas will progress down to the scale at which stars form or $\sim$0.1~pc. However, this fragmentation can start at larger scales in the cold streams or in the hot(warm) gas phase surrounding the galaxy. Some clumps of warm gas could already be formed around the galaxy. We assume in our model that a fraction of the newly accreted gas is already fragmented but we do not include any interaction between this already fragmented gas in the halo and the large scale wind. This kind of interaction could reduce, perhaps substantially, the gas-accretion efficiency \citep{Cornuault_2018}. Indeed such coupling could increase the cloud-cloud velocity dispersion and maintain the turbulence in the hot and warm gas contained in the CGM of the most massive galaxy. Such a hypothesised mechanism will be complementary to the development of thermal instabilities, could in fact act as a catalyst to the formation of more clouds in the CGM, and will could contribute in quenching the radiative cooling and dissipation in the most massive halos.

\begin{acknowledgements}
MC acknowledges Thomas Fenouillet for his much appreciated help in the use of the LAM's computation clusters and wishes to thank Mathieu G\'enois for his help in the proofreading of this article and for all physical and technical discussions. MC thanks Olivier Ilbert for the numerous and very helpfull discussion about stellar mass function (Estimations, errors, limits) and Benoit Epinat for useful discussion about galaxy disk dynamics. MC also wishes to express his appreciation to Alexandre Beelen and Yanick Roehlly for very useful physical and technical discussions. Authors thank the Centre National d'Etudes Spatiales (CNES), Aix Marseille Universit\'e, Sorbonne Universit\'e, the Programme National de Cosmologie and Galaxies (PNCG) and Programme de Physico-chimie du Milieu Interstellaire (PCMI) of CNRS/INSU for their financial support. PG gratefully acknowledges the support of the Institut Universitaire de France (IUF). 
\end{acknowledgements}

\bibliographystyle{aa} 
\bibliography{GAS_I}

\begin{appendix}

\section{The GALAKSIENN library}

The GALAKSIENN library stores the main results produced by our new \GAS\, semi-analytical model, especially MOCK galaxy catalogs and sky maps. It is available online through the ZENODO platform: {\tt{https://zenodo.org/}, DOI: 10.5281/zenodo.1451229}. A complete description of the GALAKSIENN library is given in paper III. In association with this paper I, we distribute the ASCII tables of the stellar mass functions (Fig. \ref{fig:stellar_mass_function}). 

\end{appendix}

\end{document}